\documentclass[11pt]{article} 
\usepackage[english]{babel} 
\usepackage[T1]{fontenc} 
\usepackage[utf8]{inputenc}
\usepackage{csquotes}
\usepackage{geometry} 
\usepackage{amsmath,amssymb,amsfonts,amsthm} 
\usepackage{bm} 
\usepackage{bbm} 
\usepackage{physics} 
\usepackage{graphicx} 
\usepackage{booktabs} 
\usepackage{xcolor} 
\usepackage{listings} 
\usepackage[ backend=biber, style=numeric, sorting=nyt, giveninits=true, doi=true, url=true, isbn=false ]{biblatex} 
\usepackage[ colorlinks=true, linkcolor=blue, citecolor=blue, urlcolor=blue ]{hyperref} 
\definecolor{codebackground}{rgb}{0.96,0.96,0.96} \definecolor{codegreen}{rgb}{0.00,0.45,0.00} \definecolor{codepurple}{rgb}{0.45,0.00,0.55} \definecolor{codegray}{rgb}{0.45,0.45,0.45} 
\lstdefinestyle{Rstyle}{ language=R, backgroundcolor=\color{codebackground}, basicstyle=\ttfamily\small, commentstyle=\color{codegreen}, keywordstyle=\color{blue}, stringstyle=\color{codepurple}, numberstyle=\tiny\color{codegray}, numbers=left, numbersep=8pt, stepnumber=1, breaklines=true, breakatwhitespace=false, keepspaces=true, showstringspaces=false, frame=single, captionpos=b, tabsize=2, columns=fullflexible } 
\newcommand{\indep}{\perp\!\!\!\perp}

\title{Recovering Posterior Beliefs in Credit Risk: A Latent-State EM Extension of the Information-Geometric Framework}

\author{ Lorenzo Quirini\thanks{Head of \emph{Innovazione del Credito} at Banca Monte dei Paschi di Siena. The views expressed in this paper are solely those of the author and do not represent the views, positions, or policies of the author's employing bank. Any errors are the author's own.}}

\date{September 2026}

\begin{document}

\maketitle

\begin{abstract} This paper develops a latent-state framework for recovering borrower-level posterior beliefs in credit-risk analysis. Creditworthiness and financial fragility are represented as latent dimensions, while observed borrower scores follow a finite Gaussian mixture model and default depends on the latent profile. Borrower-specific probabilities of default are obtained by averaging profile-specific default probabilities over the posterior distribution of latent states. A joint Expectation--Maximization procedure is used to estimate the mixture structure and the profile-specific default probabilities from observed score--default pairs. After estimation, predictive posterior beliefs are computed using the observed scores alone, thereby preserving the information available before default realization. A controlled simulation experiment evaluates the recovery of structural parameters, posterior beliefs, and borrower-level probabilities of default. Posterior distributions are interpreted as points on the probability simplex, and their recovery is assessed using both conventional error measures and information-geometric divergences. The results provide a controlled benchmark for studying the interaction between latent economic structure, posterior uncertainty, and credit-risk prediction. \end{abstract}


\section{Introduction} 
A previous information-geometric framework~\cite{Quirini2026} represented borrower-level credit risk through posterior distributions over possible economic states, interpreting each distribution as an information state of the financial intermediary and geometrically as a point on a statistical manifold.
The present paper takes this interpretation as its starting point and addresses the inverse statistical problem. If credit risk is represented by a posterior belief over latent borrower states, can such a belief be recovered from observed borrower-level data? More specifically, can the distribution representing the bank's uncertainty about the borrower's latent creditworthiness and financial-fragility states be estimated from observable creditworthiness and financial-fragility indicators? This question is important because the latent state of a borrower and the information available to the bank are conceptually distinct objects. The former refers to the underlying financial and economic conditions of the borrower, while the latter is represented by the posterior distribution that the intermediary assigns to the possible latent states after observing borrower-level information. 
To study this problem in a controlled setting, we develop a finite latent-state model in which creditworthiness and financial fragility are represented as unobserved borrower dimensions. Observed creditworthiness and fragility scores are generated by a Gaussian mixture model, while default is assumed to depend on the latent borrower profile. The probability of default of each borrower is then obtained by averaging profile-specific default probabilities over the posterior distribution of latent states. The purpose of the simulation experiment is to evaluate whether the posterior belief distribution itself can be recovered from observed data. Since the true data-generating process is known by construction, the estimated posterior distributions can be compared with their true counterparts. This allows us to assess the recovery of the latent population structure, the accuracy of borrower-level posterior beliefs, and the resulting probability of default. 
The connection with the previous framework~\cite{Quirini2026} is therefore direct. That framework provides the geometric interpretation of posterior credit-risk beliefs on a statistical manifold, while the present paper studies how such beliefs can be generated, estimated, and validated in a latent-variable model.
Once the posterior distributions have been recovered, they can again be interpreted as points in a statistical space. This makes it possible to evaluate the estimation procedure using information-geometric quantities such as entropy and Kullback--Leibler divergence. The contribution of the paper is threefold. 
\begin{itemize} 
\item First, it formulates credit-risk prediction as a problem of posterior belief recovery rather than as a problem of hard classification. 
\item Second, it introduces a joint latent-state EM framework that estimates both the Gaussian mixture structure and the profile-specific probabilities of default, while distinguishing the posterior profile-membership probabilities used within the EM algorithm, hereafter referred to as responsibilities, from the score-based posterior beliefs used for predictive credit-risk assessment.
\item Third, it provides an information-geometric validation of the recovered posterior beliefs by comparing true and estimated distributions on the probability simplex. 
\end{itemize}
The remainder of the paper is organized as follows. Section~2 introduces the latent-state generative model and its structural assumptions. Section~3 describes the controlled simulation design. Section~4 develops the joint EM estimation procedure. Section~5 defines the validation protocol and label-alignment rule. Section~6 reports the simulation results. Section~7 discusses their interpretation, limitations, and implications for empirical credit-risk modeling. Section~8 concludes and outlines directions for further research.

\subsection{Related Literature} The present contribution lies at the intersection of four strands of literature: credit-risk modelling, finite-mixture and latent-class models, posterior uncertainty quantification, and information geometry. The credit-risk literature traditionally represents borrower risk through a probability of default, a rating class, or a score constructed from observed borrower characteristics. Standard references discuss statistical credit-scoring methods, default prediction, portfolio applications, and the practical implementation of credit-risk models \cite{Anderson2007,Thomas2002,Thomas2009}.

Latent-variable and mixture representations provide a natural way to model unobserved borrower heterogeneity, particularly when observed default behaviour is interpreted as being driven by unobserved borrower or systematic risk factors. The present paper shares with this literature the objective of representing unobserved heterogeneity, but differs in its primary inferential target. The central object is not only the probability of default or the latent class assigned to a borrower, but the complete posterior distribution over the possible latent borrower profiles. From a statistical perspective, the proposed model belongs to the broad family of finite-mixture and model-based clustering methods \cite{FraleyRaftery2002,McLachlanPeel2000,VermuntMagidson2002}. In these approaches, population heterogeneity is represented through a finite collection of latent components, each characterized by a probability distribution for the observed variables. Posterior component-membership probabilities provide a probabilistic alternative to deterministic cluster assignments, while the Expectation--Maximization algorithm offers a natural procedure for maximum-likelihood estimation \cite{Bishpo2006,McLachlanPeel2000}. In the present framework, the observed creditworthiness and financial-fragility scores follow a finite Gaussian mixture, but the model is extended by associating each latent profile with a profile-specific probability of default. The Gaussian mixture parameters and the profile-specific default probabilities are therefore estimated jointly from observed score--default pairs. An important distinction of the proposed approach concerns the role assigned to posterior probabilities. In conventional model-based clustering, posterior component-membership probabilities are commonly used to allocate observations to latent groups, often through a modal classification rule. Such a rule retains only the most probable component and discards the remaining posterior mass. The present paper instead treats the complete score-based posterior vector as an inferential object in its own right. This distinction is substantive in credit-risk applications because two borrowers assigned to the same modal profile may have different posterior distributions and, therefore, different configurations of uncertainty concerning their latent economic conditions. The paper also distinguishes two posterior quantities that perform different statistical functions. During historical estimation, the EM algorithm uses posterior profile-membership probabilities conditional on both the observed score vector and the realized default outcome. After model estimation, the posterior distribution relevant for credit-risk assessment is recomputed using the observed scores alone. The realized default outcome therefore contributes to the estimation of the structural parameters without being used as a predictor of its own borrower-level probability of default. This separation between estimation probabilities and predictive posterior beliefs is central to the interpretation of the proposed recovery procedure. The information-geometric literature provides the final component of the framework. Information geometry represents probability distributions as points on statistical manifolds and uses geometrically meaningful quantities to compare them \cite{Rao1945, Murray1993, Amari2016}. The information-geometric framework for credit-risk monitoring developed in~\cite{Quirini2026} adopted this perspective to represent the intermediary's knowledge of borrower creditworthiness and financial fragility through posterior distributions. The present paper addresses the inverse problem left open by that framework: whether such borrower-level posterior beliefs can be estimated from observable data and how accurately their position on the probability simplex can be recovered. Accordingly, the contribution of the paper is the integration of a joint latent-state credit-risk model with an information-geometric recovery objective. 

More specifically, the paper: (i) jointly estimates the latent Gaussian mixture structure and the profile-specific probabilities of default; (ii) distinguishes the posterior profile-membership probabilities used for historical estimation from the score-based posterior beliefs used for predictive assessment; (iii) derives borrower-level probabilities of default as functionals of the recovered posterior distributions; and (iv) evaluates recovery directly on the probability simplex using Euclidean errors, Kullback--Leibler divergence, Jensen--Shannon divergence, and the information loss induced by a product-of-marginals approximation. The controlled simulation therefore provides a benchmark for assessing not only parameter and prediction recovery, but also recovery of the complete borrower-level information state.

\section{Latent-State Generative Model and Structural Assumptions}
\label{sec:latent_model}

The framework developed in \cite{Quirini2026} represents credit risk as an
information state concerning the borrower rather than as a point-valued
characteristic. In particular, the creditworthiness and financial fragility
of a borrower are represented as latent dimensions, while the information
available to the financial intermediary is summarized by a posterior
distribution over these latent states. The present paper builds on this
framework by considering the inverse problem: given observed borrower-level
data generated by an underlying latent structure, can the posterior
distribution representing the bank's state of knowledge be recovered from
the data?

\subsection{Latent creditworthiness and financial fragility}

Let borrower $i$ be characterized by two latent dimensions,

\begin{equation}
\boldsymbol{Z}_i =
\begin{pmatrix}
Z_{Ri}\\
Z_{Fi}
\end{pmatrix},
\end{equation}

where $Z_{Ri}$ denotes the latent creditworthiness state and $Z_{Fi}$
denotes the latent financial fragility state.

The two latent dimensions are not directly observable. Instead, the
financial intermediary observes a vector of indicators or scores,

\begin{equation}
\boldsymbol{X}_i =
\begin{pmatrix}
R_i\\
F_i
\end{pmatrix},
\end{equation}

where $R_i$ denotes an observed creditworthiness score and $F_i$ denotes an observed financial-fragility score.

The fundamental object of interest is therefore the conditional
distribution

\begin{equation}
p(\boldsymbol{Z}_i \mid \boldsymbol{X}_i),
\label{eq:posterior}
\end{equation}

which is interpreted as the state of knowledge of the financial
intermediary concerning the latent condition of borrower $i$ after
observing $\boldsymbol{X}_i$.

This interpretation is important because the latent state of the borrower
and the information available to the intermediary are conceptually
distinct objects. The former is represented by $\boldsymbol{Z}_i$,
whereas the latter is represented by the posterior distribution
$p(\boldsymbol{Z}_i \mid \boldsymbol{X}_i)$. 

\subsection{A finite latent-state representation}

In order to make the latent structure estimable in a controlled simulation
experiment, we consider a finite representation of the two-dimensional
latent space. Specifically, each latent dimension is represented by a
binary variable,

\begin{equation}
Z_{Ri} \in \{0,1\},
\qquad
Z_{Fi} \in \{0,1\}.
\end{equation}

The joint latent state therefore takes values in the set

\begin{equation}
\mathcal{Z}
=
\left\{
(0,0),(0,1),(1,0),(1,1)
\right\}.
\end{equation}

The four elements of $\mathcal{Z}$ represent four possible combinations
of latent creditworthiness and financial fragility. The binary
representation provides a finite latent-state representation that
allows the posterior distribution to be explicitly generated and
subsequently recovered from simulated observations.

Let

\begin{equation}
Z_i \in \{1,\ldots,K\},
\qquad K=4,
\end{equation}

denote the corresponding latent profile, and let

\begin{equation}
\pi_k = P(Z_i=k),
\qquad
\sum_{k=1}^{K}\pi_k=1
\end{equation}

denote the prior probability associated with latent profile $k$.
\footnote{With a slight abuse of notation, $Z_i$ will also denote the categorical index associated with the joint configuration $(Z_{Ri},Z_{Fi})$.}

\subsection{Gaussian mixture model for the observed information}

Conditional on the latent profile $Z_i=k$, the observed score vector is
assumed to follow a bivariate Gaussian distribution,

\begin{equation}
\boldsymbol{X}_i \mid Z_i=k
\sim
\mathcal{N}
\left(
\boldsymbol{\mu}_k,
\boldsymbol{\Sigma}_k
\right),
\qquad
k=1,\ldots,K,
\label{eq:conditional_gaussian}
\end{equation}

where $\boldsymbol{\mu}_k$ and $\boldsymbol{\Sigma}_k$ denote,
respectively, the mean vector and covariance matrix associated with
latent profile $k$.

The marginal distribution of the observed information is consequently

\begin{equation}
p(\boldsymbol{x})
=
\sum_{k=1}^{K}
\pi_k
\mathcal{N}
\left(
\boldsymbol{x};
\boldsymbol{\mu}_k,
\boldsymbol{\Sigma}_k
\right).
\label{eq:gmm}
\end{equation}

Equation~\eqref{eq:gmm} provides the generative model for the observed
borrower population. Each Gaussian component represents a region of the
observed score space associated with a particular latent combination of
creditworthiness and financial fragility.

The parameter set governing the distribution of the observed information
is

\begin{equation}
\Theta_X
=
\left\{
\pi_k,\boldsymbol{\mu}_k,\boldsymbol{\Sigma}_k
\right\}_{k=1}^{K}.
\end{equation}

In contrast with the previous framework \cite{Quirini2026}, where the parameters governing
the latent-observed relationship can be specified exogenously, the
present framework treats $\Theta_X$ as an estimable quantity. In the
simulation experiment, the parameters are first specified in the
data-generating process and are subsequently estimated from the observed
data using the expectation-maximization (EM) algorithm.

\subsection{The structural default model}

Let

\begin{equation}
D_i \in \{0,1\}
\end{equation}

denote the default outcome of borrower $i$, where $D_i=1$ indicates
default.

We associate each latent profile with a profile-specific probability of
default,

\begin{equation}
p_k
=
P(D_i=1\mid Z_i=k),
\label{eq:profile_pd}
\end{equation}

and assume

\begin{equation}
D_i\mid Z_i=k
\sim
\operatorname{Bernoulli}(p_k).
\end{equation}

The complete generative model can therefore be written as

\begin{equation}
\boxed{
p(\boldsymbol{x}_i,z_i,d_i)
=
p(z_i)
p(\boldsymbol{x}_i\mid z_i)
p(d_i\mid z_i).
}
\label{eq:joint_model}
\end{equation}

The factorization in Equation~\eqref{eq:joint_model} implies the
conditional independence assumption

\begin{equation}
\boxed{
D_i \indep \boldsymbol{X}_i
\mid Z_i.
}
\label{eq:conditional_independence}
\end{equation}

This assumption provides the structural benchmark for the controlled recovery experiment.

Conditional on the latent state of the borrower, the
observed scores do not contain additional information concerning default.

The assumption should not be interpreted as an empirical assertion that
holds in all applications. Instead, it provides a theoretically
controlled benchmark in which the relationship between observed
information, latent states, posterior beliefs, and default probability
can be studied separately. In particular, it allows borrower-level default probability to be represented as the posterior average of the structural default probabilities associated with the possible latent states.

\subsection{Posterior beliefs and the information state of the bank}

Given an observed vector $\boldsymbol{x}_i$, Bayes' theorem yields the
posterior probability that borrower $i$ belongs to latent profile $k$,

\begin{equation}
\gamma_{ik}
=
P(Z_i=k\mid\boldsymbol{X}_i=\boldsymbol{x}_i)
=
\frac{
\pi_k
\mathcal{N}
\left(
\boldsymbol{x}_i;
\boldsymbol{\mu}_k,
\boldsymbol{\Sigma}_k
\right)
}{
\displaystyle
\sum_{j=1}^{K}
\pi_j
\mathcal{N}
\left(
\boldsymbol{x}_i;
\boldsymbol{\mu}_j,
\boldsymbol{\Sigma}_j
\right)
}.
\label{eq:posterior_probability}
\end{equation}

The posterior vector

\begin{equation}
\boldsymbol{\gamma}_i
=
(\gamma_{i1},\ldots,\gamma_{iK})
\label{eq:posterior_vector}
\end{equation}

is therefore the probabilistic representation of the intermediary's
knowledge concerning the latent condition of borrower $i$.

This distinction between latent state and posterior belief is central to
the framework. The variable $Z_i$ represents the underlying state of the
borrower, while $\boldsymbol{\gamma}_i$ represents the intermediary's
uncertainty concerning that state conditional on the available
information.

Accordingly, the most probable latent profile does not provide a complete representation of the uncertainty surrounding the latent classification. The full posterior vector is required to preserve both the relative plausibility of the alternative profiles and the degree of posterior concentration.

\subsection{Default probability as a functional of posterior beliefs}

Under the structural model, the borrower-specific probability of
default conditional on the observed information can be obtained by
integrating the profile-specific probabilities of default over the
posterior distribution.
By the law of total probability and the conditional independence assumption in Equation~\eqref{eq:conditional_independence},

\begin{align}
P(D_i=1\mid\boldsymbol{X}_i=\boldsymbol{x}_i)
&=
\sum_{k=1}^{K}
P(D_i=1\mid Z_i=k)
P(Z_i=k\mid\boldsymbol{X}_i=\boldsymbol{x}_i)
\\
&=
\sum_{k=1}^{K}
p_k\gamma_{ik}.
\end{align}

Hence,
\begin{equation} \mathrm{PD}_i = \mathbb{E}_{Z_i \mid X_i} \left[ P(D_i=1\mid Z_i) \right]. 
\label{eq:pd_posterior}
\end{equation}

Equation~\eqref{eq:pd_posterior} establishes the relationship between latent-state inference and credit-risk prediction. The
probability of default is obtained by averaging the default probabilities
associated with the possible latent states according to the posterior
beliefs generated by the observed information.

The posterior distribution is therefore not merely an intermediate
quantity used for classification. It is the object through which the
information available to the intermediary is translated into a
borrower-specific probability of default.

\subsection{Product-of-marginals representation}
The posterior vector introduced in Equation~\eqref{eq:posterior_vector} assigns probabilities to the $K=4$ joint latent profiles. Because each profile corresponds to a specific configuration of latent creditworthiness and financial fragility, the same posterior distribution can equivalently be written as a joint distribution over the two binary latent dimensions: \begin{equation} p_i^{X}(z_R,z_F) = P \left( Z_{Ri}=z_R, Z_{Fi}=z_F \mid \boldsymbol{X}_i \right), \qquad z_R,z_F\in\{0,1\}. \label{eq:joint_latent_posterior} \end{equation}
Under the profile ordering \[ (0,0),\ (0,1),\ (1,0),\ (1,1), \] the four entries of this joint posterior satisfy \begin{equation} \begin{aligned} p_i^{X}(0,0) &= \gamma_{i1}, & p_i^{X}(0,1) &= \gamma_{i2}, \\ p_i^{X}(1,0) &= \gamma_{i3}, & p_i^{X}(1,1) &= \gamma_{i4}. \end{aligned} \label{eq:gamma_joint_correspondence} \end{equation} Thus, $p_i^{X}(z_R,z_F)$ and $\boldsymbol{\gamma}_i$ are two equivalent representations of the same borrower-level posterior belief.

The posterior distribution of latent creditworthiness is obtained by summing the joint posterior over the possible values of latent financial fragility: \begin{equation} p_{R,i}^{X}(z_R) = \sum_{z_F\in\{0,1\}} p_i^{X}(z_R,z_F). \label{eq:creditworthiness_posterior_marginal} \end{equation} Similarly, the posterior distribution of latent financial fragility is obtained by summing over the possible values of latent creditworthiness: \begin{equation} p_{F,i}^{X}(z_F) = \sum_{z_R\in\{0,1\}} p_i^{X}(z_R,z_F). \label{eq:fragility_posterior_marginal} \end{equation}

With the adopted profile ordering, these marginal probabilities are \begin{equation} \begin{aligned} p_{R,i}^{X}(0) &= \gamma_{i1}+\gamma_{i2}, & p_{R,i}^{X}(1) &= \gamma_{i3}+\gamma_{i4}, \\ p_{F,i}^{X}(0) &= \gamma_{i1}+\gamma_{i3}, & p_{F,i}^{X}(1) &= \gamma_{i2}+\gamma_{i4}. \end{aligned} \label{eq:posterior_marginals_from_gamma} \end{equation}

The joint score-based posterior generally contains not only marginal uncertainty about the two latent dimensions, but also posterior dependence between them. To isolate the informational role of this dependence, we construct a product-of-marginals approximation: \begin{equation} q_i^{\mathrm{PM}}(z_R,z_F) = p_{R,i}^{X}(z_R) p_{F,i}^{X}(z_F). \label{eq:product_of_marginals} \end{equation} This approximation preserves exactly the two posterior marginal distributions but removes their posterior dependence. 

Consequently, \begin{equation} q_i^{\mathrm{PM}}(z_R,z_F) = p_i^{X}(z_R,z_F) \end{equation} if and only if latent creditworthiness and latent financial fragility are conditionally independent under the score-based posterior: \begin{equation} Z_{Ri} \indep Z_{Fi} \mid \boldsymbol{X}_i. \end{equation}

This posterior-independence condition is distinct from the structural conditional-independence assumption \[ D_i \indep \boldsymbol{X}_i \mid Z_i, \] introduced in Equation~\eqref{eq:conditional_independence}. The latter concerns the data-generating relationship between scores and default, whereas the former concerns the dependence structure between the two latent dimensions under the score-based posterior distribution.

The borrower-specific probability of default implied by this representation is \begin{equation} \mathrm{PD}_{i}^{\mathrm{PM}} = \sum_{z_R} \sum_{z_F} q_i^{\mathrm{PM}}(z_R,z_F) p_{z_R,z_F},
\label{eq:pd_product_of_marginals} \end{equation}

where \[ p_{z_R,z_F} = P \left( D_i=1 \mid Z_{Ri}=z_R, Z_{Fi}=z_F \right) \] denotes the profile-specific probability of default associated with the joint latent configuration $(z_R,z_F)$.
Under the adopted profile ordering, $p_{z_R,z_F}$ corresponds to the appropriate element of the vector $(p_1,\ldots,p_K)$.

Comparing $p_i^{X}$ with $q_i^{\mathrm{PM}}$ therefore makes it possible to distinguish marginal posterior uncertainty from the additional information carried by posterior dependence. The corresponding Kullback--Leibler divergence is equal to the posterior mutual information between latent creditworthiness and latent financial fragility and is evaluated in Section~\ref{subsec:results_mean_field}.

\subsection{Information-geometric characterization}

For each borrower, the posterior vector

\begin{equation}
\boldsymbol{\gamma}_i
\end{equation}

belongs to the probability simplex

\begin{equation}
\Delta^{K-1}
=
\left\{
\boldsymbol{q}\in\mathbb{R}^{K}:
q_k\geq0,\;
\sum_{k=1}^{K}q_k=1
\right\}.
\end{equation}

The inclusion of $\boldsymbol{\gamma}_i$ in the probability simplex follows directly from its probabilistic definition. For every borrower $i$, each coordinate satisfies \begin{equation} \gamma_{ik} = P(Z_i = k \mid X_i = x_i) \geq 0, \qquad k=1,\ldots,K, \label{eq:posterior_nonnegativity} \end{equation} and, because the latent profiles are mutually exclusive and collectively exhaustive, \begin{equation} \sum_{k=1}^{K} \gamma_{ik} = \sum_{k=1}^{K} P(Z_i = k \mid X_i = x_i) = 1. \label{eq:posterior_normalization} \end{equation} Consequently, \begin{equation} \boldsymbol{\gamma}_i \in \Delta^{K-1}. \label{eq:posterior_simplex_membership} \end{equation} Although $\boldsymbol{\gamma}_i$ has $K$ coordinates and is embedded in $\mathbb{R}^{K}$, the normalization constraint leaves only $K-1$ independent coordinates. This explains why the corresponding probability simplex has dimension $K-1$. In the present model, $K=4$, so each posterior belief is represented by a point in the three-dimensional simplex $\Delta^3$ embedded in $\mathbb{R}^4$.

The vertices of the simplex correspond to degenerate posterior distributions assigning unit probability to a single latent profile. Points in the interior instead assign positive probability to multiple profiles and therefore represent different degrees and configurations of posterior uncertainty. The position of $\boldsymbol{\gamma}_i$ within the simplex consequently contains more information than the modal assignment $\arg\max_k \gamma_{ik}$, which retains only the most probable latent profile.

The posterior distributions generated by the latent-state model can
therefore be regarded as points in a statistical space. This provides
the link with the information-geometric framework developed in \cite{Quirini2026}.

In particular, posterior uncertainty can be quantified by the Shannon
entropy,

\begin{equation}
H(\boldsymbol{\gamma}_i)
=
-\sum_{k=1}^{K}
\gamma_{ik}\log\gamma_{ik}.
\label{eq:entropy}
\end{equation}

The informational separation between two posterior beliefs can be measured by the Kullback--Leibler divergence:

\begin{equation}
D_{\mathrm{KL}}
\left(
\boldsymbol{\gamma}_i
\middle\|
\boldsymbol{\gamma}_j
\right)
=
\sum_{k=1}^{K}
\gamma_{ik}
\log
\frac{\gamma_{ik}}{\gamma_{jk}}.
\label{eq:kl}
\end{equation}

These quantities allow the simulation experiment to move beyond the
recovery of latent classifications. The objective is to assess whether
the statistical procedure can recover the posterior distributions that represent the model-implied information states of the intermediary
and, consequently, the
information-geometric quantities associated with those distributions.

\subsection{Data-generating process}

The simulation experiment is based on the following generative sequence.
For each borrower $i$,

\begin{align}
Z_i
&\sim
\operatorname{Categorical}
(\pi_1,\ldots,\pi_K),
\\
\boldsymbol{X}_i\mid Z_i=k
&\sim
\mathcal{N}
(\boldsymbol{\mu}_k,\boldsymbol{\Sigma}_k),
\\
D_i\mid Z_i=k
&\sim
\operatorname{Bernoulli}(p_k).
\end{align}

The complete parameter set of the data-generating process is therefore

\begin{equation}
\Theta
=
\left\{
\pi_k,
\boldsymbol{\mu}_k,
\boldsymbol{\Sigma}_k,
p_k
\right\}_{k=1}^{K}.
\end{equation}

The parameters $\Theta$ are known by construction in the simulation
experiment. They therefore provide the benchmark against which the
estimated quantities can be evaluated.

The empirical exercise proceeds by observing only \(\{(X_i,D_i)\}_{i=1}^{n}\) and treating the latent states \(Z_i\) as unobserved. The joint EM algorithm is applied to these historical observations in order to estimate the Gaussian mixture structure and the profile-specific probabilities of default. After estimation, the borrower-level posterior beliefs used for credit-risk assessment are computed conditionally on \(X_i\) alone. The simulation consequently allows us to evaluate separately the recovery of the latent mixture structure, the recovery of the profile-specific probabilities of default, the recovery of score-based posterior beliefs, the accuracy of the resulting borrower-level probabilities of default, and the information loss associated with the product-of-marginals approximation.

\section{Baseline Simulation Experiment}
\label{sec:first_simulation}

The latent-state model introduced in Section~\ref{sec:latent_model}
provides a controlled framework in which the relationship between latent
borrower characteristics, observed scores, posterior beliefs, and default
can be studied separately. The purpose of the baseline simulation experiment
is to construct a simple data-generating process in which the latent state
is known by construction, while the bank observes only a set of borrower
scores and the subsequent default outcome.

The experiment is deliberately parsimonious. Each borrower is characterized
by two binary latent variables, representing creditworthiness and financial
fragility. The observed information is represented by two continuous scores
that, conditional on the latent state, follow a bivariate Gaussian
distribution. Default is generated from the latent state and is therefore
conditionally independent of the observed scores.

This experiment provides a benchmark for the subsequent estimation
exercise. Since the parameters of the data-generating process are known,
the posterior distribution obtained from the estimation procedure can
later be compared with the true posterior distribution implied by the
model.

\subsection{Latent borrower profiles}

We consider two binary latent variables,

\begin{equation}
Z_R \in \{0,1\},
\qquad
Z_F \in \{0,1\},
\end{equation}

where $Z_R$ denotes the latent creditworthiness state and $Z_F$ denotes
the latent financial fragility state.

The joint latent state is therefore

\begin{equation}
Z=(Z_R,Z_F),
\end{equation}

and takes four possible values,

\begin{equation}
\mathcal{Z}
=
\left\{
(0,0),(0,1),(1,0),(1,1)
\right\}.
\end{equation}

We associate each combination of the two latent dimensions with a latent
borrower profile. For convenience, the four profiles are indexed by $k=1,\ldots,4$, as shown in Table~\ref{tab:latent_profiles}.
\begin{table}[ht]
\centering
\caption{Latent profiles}
\label{tab:latent_profiles}
\begin{tabular}{c c c l}
\hline
Profile & $Z_R$ & $Z_F$ & Interpretation \\
\hline
$1$ & $0$ & $0$ & High creditworthiness / Low fragility \\
$2$ & $0$ & $1$ & High creditworthiness / High fragility \\
$3$ & $1$ & $0$ & Low creditworthiness / Low fragility \\
$4$ & $1$ & $1$ & Low creditworthiness / High fragility \\
\hline
\end{tabular}
\end{table}

The prior probabilities of the four latent profiles are denoted by $\pi_k=P(Z=k)$.
In the baseline experiment we set

\begin{equation}
\boldsymbol{\pi}
=
(0.40,0.20,0.25,0.15).
\label{eq:pi_sim}
\end{equation}

The probabilities in Equation~\eqref{eq:pi_sim} imply a heterogeneous
population in which the four latent profiles occur with different
frequencies.

\subsection{Generation of the observed scores}

For each borrower, the bank observes two continuous scores,

\begin{equation}
\boldsymbol{X}
=
\begin{pmatrix}
R\\
F
\end{pmatrix},
\end{equation}

where $R$ denotes the observed creditworthiness score and $F$ denotes the observed
financial-fragility score.
\footnote{The observed variables $R$ and $F$ are continuous synthetic scores with different economic orientations. Higher values of $R$ indicate stronger creditworthiness and therefore a more favorable credit-risk profile, whereas higher values of $F$ indicate greater financial fragility and therefore a less favorable financial condition. The interpretation concerns the ordering of the scores; their numerical origin and scale are specific to the simulation design.}

Conditional on the latent profile $Z=k$, the observed vector is generated
according to

\begin{equation}
\boldsymbol{X}\mid Z=k
\sim
\mathcal{N}
\left(
\boldsymbol{\mu}_k,
\boldsymbol{\Sigma}_k
\right).
\label{eq:sim_gaussian}
\end{equation}

The mean vectors are chosen as 
\begin{align} 
\mu_1 &= 
\begin{pmatrix} 1.8
\\ 0.3 
\end{pmatrix}, 
& 
\mu_2 &= 
\begin{pmatrix} 1.8\\ 1.8 
\end{pmatrix}, 
\\ \mu_3 &= 
\begin{pmatrix} -0.8
\\ 0.3 
\end{pmatrix}, 
& 
\mu_4 &= \begin{pmatrix} -0.8
\\ 1.8 
\end{pmatrix}. 
\end{align}

The covariance matrices are chosen as

\begin{align} \Sigma_1 &= \begin{pmatrix} 0.40 & 0.10\\ 0.10 & 0.40 \end{pmatrix}, & \Sigma_2 &= \begin{pmatrix} 0.40 & 0.10\\ 0.10 & 0.40 \end{pmatrix}, \\ \Sigma_3 &= \begin{pmatrix} 0.45 & 0.10\\ 0.10 & 0.45 \end{pmatrix}, & \Sigma_4 &= \begin{pmatrix} 0.45 & 0.10\\ 0.10 & 0.45 \end{pmatrix}. \end{align}

The Gaussian components retain a moderate degree of overlap, so that the observed score vector does not determine the latent borrower profile with certainty. At the same time, the component means and covariance matrices are chosen to avoid excessive overlap between the Gaussian components. This parameterization generates non-degenerate posterior beliefs while preserving sufficient statistical identification for the baseline recovery experiment.

The marginal distribution of the observed scores is consequently the
four-component Gaussian mixture

\begin{equation}
p(\boldsymbol{x})
=
\sum_{k=1}^{4}
\pi_k
\mathcal{N}
\left(
\boldsymbol{x};
\boldsymbol{\mu}_k,
\boldsymbol{\Sigma}_k
\right).
\label{eq:sim_mixture}
\end{equation}

\subsection{Generation of the default outcome}

Let

\begin{equation}
D\in\{0,1\}
\end{equation}

denote the default indicator.

In the baseline experiment, default is generated exclusively through the
latent state. We assign the following profile-specific probabilities of
default:

\begin{equation}
\boldsymbol{p}
=
(p_1,p_2,p_3,p_4)
=
(0.01,0.05,0.10,0.25).
\label{eq:pd_profiles}
\end{equation}

Therefore,

\begin{equation}
D\mid Z=k
\sim
\operatorname{Bernoulli}(p_k).
\label{eq:sim_default}
\end{equation}

The generative structure can thus be represented schematically as

\begin{equation}
Z
\longrightarrow
X
\end{equation}

and

\begin{equation}
Z
\longrightarrow
D.
\end{equation}

Consequently,

\begin{equation}
\boxed{
D\indep X\mid Z.
}
\label{eq:sim_cond_ind}
\end{equation}

This conditional independence assumption is particularly useful in the baseline experiment because it provides a structural benchmark for the controlled recovery exercise. Once the
latent state is known, the observed scores do not provide additional
information about default.

The probability of default conditional on the observed information is
nevertheless heterogeneous across borrowers because the observed scores
generate different posterior beliefs about the latent state.

\subsection{Implementation in R}

The simulation experiment is implemented in \texttt{R}. We generate a
population of \(N=5{,}000\) borrowers and set \(K=4\), corresponding to the
four latent borrower profiles defined by the binary dimensions of
creditworthiness and financial fragility. 

The implementation follows directly the data-generating process.

For reproducibility, the random seed is fixed before generating the simulated
population. The prior probabilities, component-specific mean vectors,
covariance matrices, and default probabilities are specified as follows.

The baseline parametrization is chosen to provide a controlled benchmark in which the four latent profiles remain sufficiently separated for reliable recovery while retaining non-negligible posterior uncertainty. The component means and within-component variances therefore balance statistical identification against intrinsic overlap in the observed score space. This allows the simulation to distinguish estimation error from ambiguity inherent in the data-generating process.

\begin{lstlisting}[
    style=Rstyle,
    caption={Specification of the data-generating parameters},
    label={lst:dgp-parameters}
]
set.seed(1234)

N <- 5000
K <- 4

# Prior probabilities of the latent borrower profiles
pi_true <- c(0.40, 0.20, 0.25, 0.15)

# Component-specific mean vectors
mu_true <- list(
  c( 1.8, 0.3),
  c( 1.8, 1.8),
  c(-0.8, 0.3),
  c(-0.8, 1.8)
)

# Component-specific covariance matrices
Sigma_true <- list(
  matrix(c(0.40, 0.10,
           0.10, 0.40), nrow = 2, byrow = TRUE),
  matrix(c(0.40, 0.10,
           0.10, 0.40), nrow = 2, byrow = TRUE),
  matrix(c(0.45, 0.10,
           0.10, 0.45), nrow = 2, byrow = TRUE),
  matrix(c(0.45, 0.10,
           0.10, 0.45), nrow = 2, byrow = TRUE)
)

# Profile-specific default probabilities
pd_true <- c(0.01, 0.05, 0.10, 0.25)
\end{lstlisting}

The first step consists of generating the latent profile of each borrower.
Since the latent state is unobserved in empirical applications, this variable
is used here only because the simulation experiment provides access to the
true data-generating structure.

\begin{lstlisting}[
    style=Rstyle,
    caption={Generation of latent borrower profiles},
    label={lst:latent-profiles}
]
Z <- sample(
  x = 1:K,
  size = N,
  replace = TRUE,
  prob = pi_true
)
\end{lstlisting}

The profile index \(Z_i\) can then be mapped into the two binary latent
dimensions \(Z_{Ri}\) and \(Z_{Fi}\). These variables represent, respectively,
the latent creditworthiness state and the latent financial-fragility state.

\begin{lstlisting}[
    style=Rstyle,
    caption={Mapping latent profiles into binary latent dimensions},
    label={lst:binary-latent-dimensions}
]
profiles <- data.frame(
  Z_R = c(0, 0, 1, 1),
  Z_F = c(0, 1, 0, 1)
)

Z_R <- profiles$Z_R[Z]
Z_F <- profiles$Z_F[Z]
\end{lstlisting}

Conditional on the latent profile, the observed score vector
\(X_i=(R_i,F_i)^\top\) is generated from a bivariate Gaussian distribution.
The following function generates multivariate Gaussian observations using the
Cholesky decomposition of the covariance matrix.

\begin{lstlisting}[
    style=Rstyle,
    caption={Function for generating bivariate Gaussian observations},
    label={lst:rmvnorm-function}
]
rmvnorm2 <- function(n, mu, Sigma) {

  L <- chol(Sigma)

  Z_standard <- matrix(
    rnorm(2 * n),
    ncol = 2
  )

  X <- Z_standard %*% L

  X <- sweep(
    X,
    MARGIN = 2,
    STATS = mu,
    FUN = "+"
  )

  return(X)
}
\end{lstlisting}

The observed score matrix is then generated component by component. Borrowers
assigned to the same latent profile are drawn from the corresponding Gaussian
component.

\begin{lstlisting}[
    style=Rstyle,
    caption={Generation of observed creditworthiness and financial-fragility scores},
    label={lst:observed-scores}
]
X <- matrix(
  NA,
  nrow = N,
  ncol = 2
)

for (k in 1:K) {

  idx <- which(Z == k)

  X[idx, ] <- rmvnorm2(
    n = length(idx),
    mu = mu_true[[k]],
    Sigma = Sigma_true[[k]]
  )
}

colnames(X) <- c("R", "F")
\end{lstlisting}

Finally, the default indicator is generated conditionally on the latent
profile. In accordance with the structural assumption of the model, default
depends on \(Z_i\) and not directly on the observed score vector \(X_i\).
Formally, the data-generating process imposes the conditional independence
assumption
\[
D_i \indep X_i \mid Z_i.
\]

\begin{lstlisting}[
    style=Rstyle,
    caption={Generation of the default outcome},
    label={lst:default-generation}
]
D <- rbinom(
  n = N,
  size = 1,
  prob = pd_true[Z]
)
\end{lstlisting}

The simulated data set combines the variables observed by the bank with the
latent variables available only in the controlled simulation environment.

\begin{lstlisting}[
    style=Rstyle,
    caption={Construction of the simulated data set},
    label={lst:simulated-dataset}
]
sim_data <- data.frame(
  R = X[, 1],
  F = X[, 2],
  Z = Z,
  Z_R = Z_R,
  Z_F = Z_F,
  D = D
)

head(sim_data)
\end{lstlisting}

In an empirical application, only \(R_i\), \(F_i\), and, when available, the
realized default indicator \(D_i\) would be observed. The latent variables
\(Z_i\), \(Z_{Ri}\), and \(Z_{Fi}\) are retained in the simulated data set
exclusively for validation purposes. Their availability allows the estimated
latent structure and the estimated posterior beliefs to be compared with the
true latent configuration generated by the model.
The code listings in this section provide a transparent implementation of the data-generating mechanism. The accompanying ancillary script implements the same calculations using log-scale Gaussian densities and log-sum-exp normalization for improved numerical stability, and contains the complete joint EM estimation and validation pipeline.

\subsection{The true posterior distribution}

The simulation also makes it possible to calculate the posterior
distribution implied by the true data-generating process. 
\footnote{Here and throughout the paper, the term ``true posterior'' denotes the posterior distribution implied by the known data-generating parameters.}

For an observed
score vector $\boldsymbol{x}_i$, Bayes' theorem gives

\begin{equation} \gamma_{ik}^{X,\ast} = P_{\Theta^\ast} \left( Z_i=k \mid X_i=x_i \right) = \frac{ \pi_k^\ast\, \mathcal{N} \left( x_i;\mu_k^\ast,\Sigma_k^\ast \right) }{ \displaystyle \sum_{j=1}^{K} \pi_j^\ast\, \mathcal{N} \left( x_i;\mu_j^\ast,\Sigma_j^\ast \right) }. \label{eq:true-score-posterior} \end{equation}

Hereafter the superscript \(\ast\) refers to the "true" values for the quantities involved in the experiment.
Accordingly, it indicates that the posterior distribution is computed using the true parameters of the data-generating process.

The superscript \(X\) emphasizes that the posterior belief is conditional only on the observed score vector and does not use the realized default outcome.

For completeness, the bivariate Gaussian density can be implemented
directly in R as follows.

\begin{lstlisting}[style=Rstyle, 
caption={Bivariate Gaussian density}, 
label={lst:dmvnorm}
]
dmvnorm2 <- function(x, mu, Sigma) {
  p <- length(mu)
  diff <- x - mu
  invS <- solve(Sigma)
  density <- (2 * pi)^(-p / 2) *
             det(Sigma)^(-1 / 2) *
             exp(-0.5 * t(diff) %*% invS %*% diff)
  return(as.numeric(density))
}
\end{lstlisting}

The true posterior probabilities can then be computed for every borrower.

\begin{lstlisting}[style=Rstyle, 
caption={Computation of the true posterior probabilities}, label={lst:true_posterior}
]
posterior_true <- matrix(
  NA,
  nrow = N,
  ncol = K
)
for (i in 1:N) {
  likelihood <- sapply(
    1:K,
    function(k) {
      dmvnorm2(
        X[i, ],
        mu_true[[k]],
        Sigma_true[[k]]
      )
    }
  )
  numerator <- pi_true * likelihood
  posterior_true[i, ] <-
    numerator / sum(numerator)
}
colnames(posterior_true) <-
  paste0("Posterior_", 1:K)
\end{lstlisting}

The matrix $\boldsymbol{\Gamma}^{*}$ generated by this procedure is a
particularly important object for the subsequent analysis. 
Its $i$-th row represents the score-based posterior distribution implied by the true data-generating parameters for borrower $i$.
In the controlled simulation, this distribution provides the benchmark for evaluating posterior recovery.

\subsection{The true probability of default conditional on observed
information}

Although default is generated conditionally only on the latent state, the
probability of default conditional on the observed scores is obtained by
integrating over the posterior distribution,

\begin{equation} PD_i^\ast = P_{\Theta^\ast}(D_i=1\mid X_i) = \sum_{k=1}^{K} p_k^\ast\, \gamma_{ik}^{X,\ast}. \label{eq:true-borrower-pd} \end{equation}

Thus, even though the structural probabilities $p_k$ are constant within
each latent profile, the conditional probability of default varies across
borrowers because their observed scores imply different posterior
distributions over the latent states.

In R, the corresponding borrower-level probability of default is obtained
as follows.

\begin{lstlisting}[style=Rstyle, 
caption={True posterior probability of default}, label={lst:true_pd}]
PD_true <- as.vector(
  posterior_true %*% pd_true
)
sim_data$PD_true <- PD_true
\end{lstlisting}

Equation~\eqref{eq:true-borrower-pd} is the key bridge between the latent-state
model and credit-risk prediction. The borrower-specific probability of
default is a functional of the posterior distribution representing the
bank's knowledge.

\subsection{Purpose of the baseline experiment}

The baseline experiment establishes a controlled benchmark for the estimation exercise developed in the following sections. The simulation provides five benchmark objects that can be compared with their estimated counterparts:

\begin{enumerate} 
\item the true latent-profile probabilities \(\pi^\ast\); 
\item the true Gaussian parameters \((\mu_k^\ast,\Sigma_k^\ast)\); 
\item the true profile-specific probabilities of default \(p_k^\ast\); 
\item the true borrower-level score-based posterior distributions \(\gamma_i^{X,\ast}\); \item the true borrower-level probabilities of default \(PD_i^\ast\). 
\end{enumerate}

The subsequent estimation exercise treats the latent states as unobserved and uses the historical pairs \((X_i,D_i)\) to estimate the Gaussian mixture structure and the profile-specific probabilities of default. After convergence, the score-based posterior beliefs are reconstructed using \(X_i\) alone. The comparison between estimated and true quantities therefore makes it possible to evaluate mixture recovery, structural default-probability recovery, posterior belief recovery, and borrower-level PD recovery separately.

\section{Joint Estimation by Expectation--Maximization}
\label{sec:em}

\subsection{Latent-variable formulation}
\label{subsec:em_latent_formulation}

Let
\begin{equation}
X_i=(R_i,F_i)^\top,
\qquad i=1,\ldots,n,
\label{eq:em_observed_scores}
\end{equation}
denote the vector of observed scores for borrower \(i\), where \(R_i\) and
\(F_i\) represent the observed creditworthiness and financial-fragility scores, respectively. Let
\begin{equation}
D_i\in\{0,1\}
\label{eq:em_observed_default}
\end{equation}
denote the observed default outcome, and let
\begin{equation}
Z_i\in\{1,\ldots,K\}
\label{eq:em_latent_profile}
\end{equation}
denote the unobserved latent borrower profile.

The model introduced in Section~\ref{sec:latent_model} assumes
\begin{equation}
P(Z_i=k)=\pi_k,
\qquad
\sum_{k=1}^{K}\pi_k=1,
\label{eq:em_mixture_probabilities}
\end{equation}
and
\begin{equation}
X_i\mid Z_i=k
\sim
\mathcal{N}_2(\mu_k,\Sigma_k).
\label{eq:em_gaussian_component}
\end{equation}

Conditional on the latent profile, the default outcome follows
\begin{equation}
D_i\mid Z_i=k
\sim
\operatorname{Bernoulli}(p_k),
\label{eq:em_bernoulli_component}
\end{equation}
where
\begin{equation}
p_k=P(D_i=1\mid Z_i=k)
\label{eq:em_profile_specific_pd}
\end{equation}
is the profile-specific probability of default.

Under the conditional independence assumption
\begin{equation}
D_i\indep X_i\mid Z_i,
\label{eq:em_conditional_independence}
\end{equation}
the joint conditional distribution of the observed variables is
\begin{equation}
p(x_i,d_i\mid Z_i=k)
=
\phi_2(x_i;\mu_k,\Sigma_k)
p_k^{d_i}(1-p_k)^{1-d_i},
\label{eq:em_joint_component_density}
\end{equation}
where \(\phi_2(\cdot;\mu_k,\Sigma_k)\) denotes the bivariate Gaussian
density.

The complete parameter set is therefore
\begin{equation}
\Theta
=
\left\{
\pi_k,\mu_k,\Sigma_k,p_k
\right\}_{k=1}^{K}.
\label{eq:em_complete_parameter_set}
\end{equation}

This formulation extends the estimation problem beyond the recovery of the
Gaussian mixture structure. The profile-specific default probabilities are
treated as unknown structural parameters and are estimated jointly with the
mixture probabilities, component means, and covariance matrices.

\subsection{Observed-data likelihood}
\label{subsec:em_observed_likelihood}

Since the latent profile \(Z_i\) is unobserved, the joint marginal distribution
of the observed score vector and default outcome is
\begin{equation}
p(x_i,d_i;\Theta)
=
\sum_{k=1}^{K}
\pi_k
\phi_2(x_i;\mu_k,\Sigma_k)
p_k^{d_i}(1-p_k)^{1-d_i}.
\label{eq:em_joint_marginal_density}
\end{equation}

The observed-data likelihood is consequently
\begin{equation}
L(\Theta;X,D)
=
\prod_{i=1}^{n}
\left[
\sum_{k=1}^{K}
\pi_k
\phi_2(X_i;\mu_k,\Sigma_k)
p_k^{D_i}(1-p_k)^{1-D_i}
\right],
\label{eq:em_joint_observed_likelihood}
\end{equation}
and the corresponding log-likelihood is
\begin{equation}
\ell(\Theta;X,D)
=
\sum_{i=1}^{n}
\log
\left[
\sum_{k=1}^{K}
\pi_k
\phi_2(X_i;\mu_k,\Sigma_k)
p_k^{D_i}(1-p_k)^{1-D_i}
\right].
\label{eq:em_joint_observed_loglikelihood}
\end{equation}

The likelihood in Equation~\eqref{eq:em_joint_observed_likelihood} uses both
the observed score vectors and the realized default outcomes. It therefore
allows the profile-specific probabilities of default to be inferred even
though the latent profile assignments are not observed.

\subsection{Complete-data log-likelihood}
\label{subsec:em_complete_likelihood}

Introduce the latent indicator variable
\begin{equation}
Y_{ik}
=
\mathbf{1}\{Z_i=k\},
\qquad
i=1,\ldots,n,
\quad
k=1,\ldots,K.
\label{eq:em_latent_indicator}
\end{equation}

If the latent indicators were observed, the complete-data likelihood would be
\begin{equation}
L_c(\Theta;X,D,Z)
=
\prod_{i=1}^{n}
\prod_{k=1}^{K}
\left[
\pi_k
\phi_2(X_i;\mu_k,\Sigma_k)
p_k^{D_i}(1-p_k)^{1-D_i}
\right]^{Y_{ik}}.
\label{eq:em_complete_likelihood}
\end{equation}

The corresponding complete-data log-likelihood is
\begin{align}
\ell_c(\Theta;X,D,Z)
={}&
\sum_{i=1}^{n}
\sum_{k=1}^{K}
Y_{ik}
\Big[
\log\pi_k
+
\log\phi_2(X_i;\mu_k,\Sigma_k)
\nonumber\\
&\qquad\qquad
+
D_i\log p_k
+
(1-D_i)\log(1-p_k)
\Big].
\label{eq:em_complete_loglikelihood}
\end{align}

The final two terms in Equation~\eqref{eq:em_complete_loglikelihood} represent
the contribution of the default outcome to the complete-data likelihood. They
are the terms from which the EM update for \(p_k\) is derived.

\subsection{E-step: posterior responsibilities conditional on scores and default}
\label{subsec:em_estep}

Given the current parameter estimate $\boldsymbol{\Theta}^{(t)}$, the E-step computes the posterior profile-membership probabilities, commonly referred to as responsibilities:

\begin{equation}
\tau_{ik}^{(t)}
=
\mathbb{E}_{\Theta^{(t)}}
\left[
Y_{ik}\mid X_i,D_i
\right]
=
P_{\Theta^{(t)}}(Z_i=k\mid X_i,D_i).
\label{eq:em_joint_responsibility_definition}
\end{equation}

By Bayes' rule,
\begin{equation}
\tau_{ik}^{(t)}
=
\frac{
\pi_k^{(t)}
\phi_2
\left(
X_i;
\mu_k^{(t)},
\Sigma_k^{(t)}
\right)
\left(p_k^{(t)}\right)^{D_i}
\left(1-p_k^{(t)}\right)^{1-D_i}
}{
\displaystyle
\sum_{j=1}^{K}
\pi_j^{(t)}
\phi_2
\left(
X_i;
\mu_j^{(t)},
\Sigma_j^{(t)}
\right)
\left(p_j^{(t)}\right)^{D_i}
\left(1-p_j^{(t)}\right)^{1-D_i}
}.
\label{eq:em_joint_responsibility}
\end{equation}

For every observation \(i\), the responsibilities satisfy
\begin{equation}
\tau_{ik}^{(t)}\geq 0,
\qquad
\sum_{k=1}^{K}\tau_{ik}^{(t)}=1.
\label{eq:em_joint_responsibility_constraints}
\end{equation}

The notation \(\tau_{ik}^{(t)}\) is used here to distinguish the posterior
responsibilities employed during joint model estimation from the posterior
beliefs based only on the observed score vector, which will be denoted by
\(\gamma_{ik}\).

\subsection{M-step: joint parameter updates}
\label{subsec:em_mstep}

Define the effective number of observations assigned to latent profile \(k\) at
iteration \(t\) as
\begin{equation}
N_k^{(t)}
=
\sum_{i=1}^{n}
\tau_{ik}^{(t)}.
\label{eq:em_joint_effective_count}
\end{equation}

The M-step maximizes the expected complete-data log-likelihood with respect to
the parameters. The updated mixture probability is
\begin{equation}
\pi_k^{(t+1)}
=
\frac{N_k^{(t)}}{n}
=
\frac{1}{n}
\sum_{i=1}^{n}
\tau_{ik}^{(t)}.
\label{eq:em_joint_update_pi}
\end{equation}

The updated component mean is
\begin{equation}
\mu_k^{(t+1)}
=
\frac{
\displaystyle
\sum_{i=1}^{n}
\tau_{ik}^{(t)}X_i
}{
N_k^{(t)}
}.
\label{eq:em_joint_update_mu}
\end{equation}

The updated covariance matrix is
\begin{equation}
\Sigma_k^{(t+1)}
=
\frac{1}{N_k^{(t)}}
\sum_{i=1}^{n}
\tau_{ik}^{(t)}
\left(
X_i-\mu_k^{(t+1)}
\right)
\left(
X_i-\mu_k^{(t+1)}
\right)^\top.
\label{eq:em_joint_update_sigma}
\end{equation}

Finally, the profile-specific probability of default is updated as
\begin{equation}
p_k^{(t+1)}
=
\frac{
\displaystyle
\sum_{i=1}^{n}
\tau_{ik}^{(t)}D_i
}{
N_k^{(t)}
}
=
\frac{
\displaystyle
\sum_{i=1}^{n}
\tau_{ik}^{(t)}D_i
}{
\displaystyle
\sum_{i=1}^{n}
\tau_{ik}^{(t)}
}.
\label{eq:em_update_profile_pd}
\end{equation}

Equation~\eqref{eq:em_update_profile_pd} has a direct interpretation. Each
observed default contributes to the estimate of \(p_k\) according to the
posterior probability that the corresponding borrower belongs to latent
profile \(k\). If the latent profiles were observed, the responsibilities
would reduce to binary indicators and Equation~\eqref{eq:em_update_profile_pd}
would coincide with the empirical default rate within profile \(k\).

\subsection{Convergence and numerical implementation}
\label{subsec:em_convergence}

At every iteration, the EM algorithm produces an updated parameter vector
\begin{equation}
\Theta^{(t+1)}
=
\left\{
\pi_k^{(t+1)},
\mu_k^{(t+1)},
\Sigma_k^{(t+1)},
p_k^{(t+1)}
\right\}_{k=1}^{K}.
\label{eq:em_updated_parameter_vector}
\end{equation}

The algorithm is stopped when the relative change in the observed-data log-likelihood satisfies 
\begin{equation} \frac{ \left| \ell(\boldsymbol{\Theta}^{(t+1)};\boldsymbol{X},\boldsymbol{D}) - \ell(\boldsymbol{\Theta}^{(t)};\boldsymbol{X},\boldsymbol{D}) \right| }{ 1+ \left| \ell(\boldsymbol{\Theta}^{(t)};\boldsymbol{X},\boldsymbol{D}) \right| } < \varepsilon.
\end{equation}

Since finite-mixture likelihoods may contain multiple local maxima, the
estimation procedure is initialized from several different parameter
configurations. Among the convergent runs, the retained solution is the one associated with the highest observed-data log-likelihood.

To avoid numerical instability, estimated profile-specific probabilities may
be constrained to the interval
\begin{equation}
p_k\in[\varepsilon_p,1-\varepsilon_p],
\label{eq:em_pd_numerical_bounds}
\end{equation}
where \(\varepsilon_p>0\) is a small numerical constant.

\subsection{Estimated structural parameters}
\label{subsec:em_estimated_structural_parameters}

Let
\begin{equation}
\widehat{\Theta}
=
\left\{
\widehat{\pi}_k,
\widehat{\mu}_k,
\widehat{\Sigma}_k,
\widehat{p}_k
\right\}_{k=1}^{K}
\label{eq:em_final_estimates}
\end{equation}
denote the parameter estimates obtained at convergence.

Since the true parameter values
\begin{equation}
\Theta^\ast
=
\left\{
\pi_k^\ast,
\mu_k^\ast,
\Sigma_k^\ast,
p_k^\ast
\right\}_{k=1}^{K}
\label{eq:em_true_parameter_set}
\end{equation}
are known in the simulation experiment, the estimation procedure can be
evaluated by comparing
\begin{equation}
\widehat{p}_k
\quad\text{with}\quad
p_k^\ast,
\qquad k=1,\ldots,K.
\label{eq:em_pd_parameter_comparison}
\end{equation}

A summary measure of profile-specific default-probability recovery is
\begin{equation}
\operatorname{MAE}_{p}
=
\frac{1}{K}
\sum_{k=1}^{K}
\left|
\widehat{p}_k-p_k^\ast
\right|,
\label{eq:em_profile_pd_mae}
\end{equation}
while the corresponding mean squared error is
\begin{equation}
\operatorname{MSE}_{p}
=
\frac{1}{K}
\sum_{k=1}^{K}
\left(
\widehat{p}_k-p_k^\ast
\right)^2.
\label{eq:em_profile_pd_mse}
\end{equation}

These measures complement the comparison of the mixture probabilities,
component means, and covariance matrices. They make it possible to assess
whether the estimation procedure recovers not only the latent distribution of
the observed scores but also the structural relationship between latent
borrower profiles and default.

\subsection{Posterior beliefs based on observed scores}
\label{subsec:em_score_posterior}

The posterior responsibilities in Equation~\eqref{eq:em_joint_responsibility}
are conditional on both \(X_i\) and \(D_i\). They are used during model
estimation because the realized default outcomes contain information about the
unknown latent profiles and the profile-specific default probabilities.

The use of \(D_i\) in the estimation responsibilities does not imply that the realized outcome is used as a predictor of its own borrower-level probability of default. The default outcome contributes only to historical parameter estimation, whereas predicted posterior beliefs and probabilities of default are computed conditionally on \(X_i\) alone.

For credit-risk prediction, however, the default outcome is not yet observed.
The relevant borrower-level posterior belief must therefore be based only on
the observed score vector. After the model has been estimated, define
\begin{equation}
\widehat{\gamma}_{ik}^{X}
=
P_{\widehat{\Theta}}(Z_i=k\mid X_i)
\label{eq:em_score_posterior_definition}
\end{equation}
as
\begin{equation}
\widehat{\gamma}_{ik}^{X}
=
\frac{
\widehat{\pi}_k
\phi_2
\left(
X_i;
\widehat{\mu}_k,
\widehat{\Sigma}_k
\right)
}{
\displaystyle
\sum_{j=1}^{K}
\widehat{\pi}_j
\phi_2
\left(
X_i;
\widehat{\mu}_j,
\widehat{\Sigma}_j
\right)
}.
\label{eq:em_score_posterior}
\end{equation}

For each borrower, the estimated posterior belief vector is
\begin{equation}
\widehat{\gamma}_i^{X}
=
\left(
\widehat{\gamma}_{i1}^{X},
\ldots,
\widehat{\gamma}_{iK}^{X}
\right).
\label{eq:em_score_posterior_vector}
\end{equation}

This posterior distribution represents the estimated information state of the
financial intermediary before the realization of the default outcome. It is
therefore the appropriate posterior distribution for borrower-level credit-risk
prediction and for the information-geometric comparisons developed in the
subsequent sections.

\subsection{Borrower-level probability of default}
\label{subsec:em_borrower_pd}

Using the estimated profile-specific probabilities of default and the posterior
beliefs based on the observed score vector, the borrower-level probability of
default is
\begin{equation}
\widehat{PD}_i
=
P_{\widehat{\Theta}}(D_i=1\mid X_i)
=
\sum_{k=1}^{K}
\widehat{p}_k
\widehat{\gamma}_{ik}^{X}.
\label{eq:em_estimated_borrower_pd}
\end{equation}

In the simulation experiment, the corresponding true probability is
\begin{equation}
PD_i^\ast
=
P_{\Theta^\ast}(D_i=1\mid X_i)
=
\sum_{k=1}^{K}
p_k^\ast
\gamma_{ik}^{X,\ast},
\label{eq:em_true_borrower_pd}
\end{equation}
where
\begin{equation}
\gamma_{ik}^{X,\ast}
=
P_{\Theta^\ast}(Z_i=k\mid X_i)
\label{eq:em_true_score_posterior}
\end{equation}
is the score-based posterior implied by the true data-generating parameters.

The difference between
\(\widehat{PD}_i\) and \(PD_i^\ast\) reflects two sources of estimation error:
the error in recovering the posterior distribution over latent states and the
error in recovering the profile-specific probabilities of default.

The borrower-level prediction error can be summarized by
\begin{equation}
\operatorname{MSE}_{PD}
=
\frac{1}{n}
\sum_{i=1}^{n}
\left(
\widehat{PD}_i-PD_i^\ast
\right)^2
\label{eq:em_borrower_pd_mse}
\end{equation}
and
\begin{equation}
\operatorname{MAE}_{PD}
=
\frac{1}{n}
\sum_{i=1}^{n}
\left|
\widehat{PD}_i-PD_i^\ast
\right|.
\label{eq:em_borrower_pd_mae}
\end{equation}

This decomposition preserves the central interpretation of the model. The
probability of default is not assigned directly to the observed score vector.
It is obtained by integrating the estimated structural default probabilities
over the posterior distribution representing the intermediary's uncertainty
about the latent borrower state.

\section{EM Recovery Experiment on the Simulated Population}
\label{sec:em_recovery}

The simulation experiment introduced in Section~\ref{sec:first_simulation} provides
a controlled environment in which the true latent states, the true Gaussian
mixture parameters, the true profile-specific probabilities of default, the
true score-based posterior beliefs, and the true borrower-level probabilities
of default are known by construction.

The purpose of the present section is to define the validation design used to
evaluate the joint EM procedure developed in Section~\ref{sec:em}. The same
simulated population generated in Section~\ref{sec:first_simulation} is used for the
estimation exercise, but the latent borrower profiles are treated as unobserved.
The estimation is therefore based only on the historical borrower-level
observations
\begin{equation}
\left\{
(X_i,D_i)
\right\}_{i=1}^{n},
\label{eq:recovery_observed_sample}
\end{equation}
where
\begin{equation}
X_i=(R_i,F_i)^\top
\label{eq:recovery_observed_scores}
\end{equation}
is the observed score vector and \(D_i\) is the realized default indicator.

The simulated latent profiles \(Z_i\), the true parameter set
\(\Theta^\ast\), the true score-based posterior distributions
\(\gamma_i^{X,\ast}\), and the true borrower-level probabilities of default
\(PD_i^\ast\) are not used as inputs to the estimation procedure. They are
retained exclusively as validation benchmarks.

This design reproduces the information structure of a historical credit-risk
sample. Borrower scores and realized default outcomes are observed, whereas the
underlying latent economic profiles are not.

\subsection{Estimation protocol}
\label{subsec:recovery_estimation_protocol}

The number of latent profiles is assumed to be known and is set equal to the
number used in the data-generating process:
\begin{equation}
K=4.
\label{eq:recovery_number_components}
\end{equation}

The complete parameter set is
\begin{equation}
\Theta
=
\left\{
\pi_k,\mu_k,\Sigma_k,p_k
\right\}_{k=1}^{K},
\label{eq:recovery_complete_parameter_set}
\end{equation}
where:

\begin{itemize}
    \item \(\pi_k\) is the population probability of latent profile \(k\);
    \item \(\mu_k\) is the profile-specific mean vector of the observed scores;
    \item \(\Sigma_k\) is the profile-specific covariance matrix;
    \item \(p_k=P(D_i=1\mid Z_i=k)\) is the profile-specific probability
    of default.
\end{itemize}

The joint EM algorithm described in Section~\ref{sec:em} is applied to the
historical observations in Equation~\eqref{eq:recovery_observed_sample}. At
convergence, the procedure produces the parameter estimates
\begin{equation}
\widehat{\Theta}
=
\left\{
\widehat{\pi}_k,
\widehat{\mu}_k,
\widehat{\Sigma}_k,
\widehat{p}_k
\right\}_{k=1}^{K}.
\label{eq:recovery_estimated_parameter_set}
\end{equation}

During estimation, the E-step computes the joint posterior responsibilities
\begin{equation}
\widehat{\tau}_{ik}
=
P_{\widehat{\Theta}}
\left(
Z_i=k\mid X_i,D_i
\right).
\label{eq:recovery_joint_responsibilities}
\end{equation}

These quantities use both the observed score vector and the realized default
outcome. They are internal objects of the estimation procedure and determine
the responsibility-weighted updates of the mixture parameters and the
profile-specific probabilities of default.

In particular, the estimated default probability associated with latent
profile \(k\) is
\begin{equation}
\widehat{p}_k
=
\frac{
\displaystyle
\sum_{i=1}^{n}
\widehat{\tau}_{ik}D_i
}{
\displaystyle
\sum_{i=1}^{n}
\widehat{\tau}_{ik}
}.
\label{eq:recovery_estimated_profile_pd}
\end{equation}

Equation~\eqref{eq:recovery_estimated_profile_pd} shows that the realized
defaults are allocated probabilistically across latent profiles. Each
observation contributes to the estimation of \(\widehat p_k\) according to
the estimated probability that the corresponding borrower belongs to profile
\(k\).

To reduce sensitivity to local maxima, the EM procedure is initialized from
multiple parameter configurations. The retained solution is the one attaining
the highest value of the joint observed-data log-likelihood
\begin{equation}
\ell(\widehat{\Theta};X,D).
\label{eq:recovery_selected_loglikelihood}
\end{equation}

In the baseline experiment, the algorithm is run from 20 initial configurations. Each initialization is obtained from a $K$-means partition with five internal starts, followed by a random perturbation of three percent of the initial labels. The maximum number of EM iterations is 300 and the relative log-likelihood tolerance is set to $10^{-8}$. A ridge term equal to $10^{-6}$ is added to the diagonal of each covariance estimate, while the profile-specific default probabilities are constrained to $[10^{-6},1-10^{-6}]$. Runs containing a component with an effective sample size smaller than 10 are discarded. Among the convergent runs, the solution with the largest observed-data log-likelihood is retained.

\subsection{Score-based posterior beliefs}
\label{subsec:recovery_score_posteriors}

After the joint model has been estimated, the borrower-level information state
used for credit-risk assessment is computed conditionally on the
observed score vector alone.

For borrower \(i\), the estimated score-based posterior probability of latent
profile \(k\) is
\begin{equation}
\widehat{\gamma}_{ik}^{X}
=
P_{\widehat{\Theta}}
\left(
Z_i=k\mid X_i
\right)
=
\frac{
\widehat{\pi}_k
\phi_2
\left(
X_i;
\widehat{\mu}_k,
\widehat{\Sigma}_k
\right)
}{
\displaystyle
\sum_{j=1}^{K}
\widehat{\pi}_j
\phi_2
\left(
X_i;
\widehat{\mu}_j,
\widehat{\Sigma}_j
\right)
}.
\label{eq:recovery_score_posterior}
\end{equation}

The corresponding posterior belief vector is
\begin{equation}
\widehat{\gamma}_i^{X}
=
\left(
\widehat{\gamma}_{i1}^{X},
\ldots,
\widehat{\gamma}_{iK}^{X}
\right).
\label{eq:recovery_score_posterior_vector}
\end{equation}

The distinction between
\(\widehat{\tau}_{ik}\) and \(\widehat{\gamma}_{ik}^{X}\) is central to the
validation design. The former is conditional on both \(X_i\) and \(D_i\) and
is used only for historical parameter estimation. The latter is conditional
on \(X_i\) alone and represents the information available before default
realization.

The use of \(D_i\) in the joint estimation responsibilities does not imply
that the realized outcome is used as a predictor of its own probability of
default. The default outcome contributes only to the estimation of the
structural parameters from historical data. Predictive posterior beliefs and
borrower-level probabilities of default are subsequently computed using only
the observed score vector.

\subsection{Borrower-level probabilities of default}
\label{subsec:recovery_borrower_pd}

The estimated borrower-level probability of default is obtained by integrating
the estimated profile-specific probabilities of default over the estimated
score-based posterior distribution:
\begin{equation}
\widehat{PD}_i
=
P_{\widehat{\Theta}}
\left(
D_i=1\mid X_i
\right)
=
\sum_{k=1}^{K}
\widehat{p}_k
\widehat{\gamma}_{ik}^{X}.
\label{eq:recovery_estimated_borrower_pd}
\end{equation}

The corresponding population benchmark is
\begin{equation}
PD_i^\ast
=
P_{\Theta^\ast}
\left(
D_i=1\mid X_i
\right)
=
\sum_{k=1}^{K}
p_k^\ast
\gamma_{ik}^{X,\ast},
\label{eq:recovery_true_borrower_pd}
\end{equation}
where
\begin{equation}
\gamma_{ik}^{X,\ast}
=
P_{\Theta^\ast}
\left(
Z_i=k\mid X_i
\right)
\label{eq:recovery_true_score_posterior}
\end{equation}
is the true score-based posterior implied by the data-generating process.

Equation~\eqref{eq:recovery_estimated_borrower_pd} uses only estimated
quantities. Consequently, the discrepancy between
\(\widehat{PD}_i\) and \(PD_i^\ast\) may arise from two sources:

\begin{enumerate}
    \item error in estimating the profile-specific probabilities of default
    \(\widehat p_k\);
    \item error in recovering the score-based posterior beliefs
    \(\widehat\gamma_{ik}^{X}\).
\end{enumerate}

The simulation design makes it possible to evaluate these two error channels
separately, since \(p_k^\ast\) and \(\gamma_{ik}^{X,\ast}\) are both known by
construction.

\subsection{Alignment of estimated and true latent profiles}
\label{subsec:recovery_label_alignment}

The likelihood of a finite mixture model is invariant with respect to
permutations of the component labels. As a consequence, the numerical labels
assigned by the EM algorithm need not coincide with the labels used in the
data-generating process.

For example, the component identified numerically as the first estimated
component may correspond economically to the third true latent profile.
Component-specific comparisons are therefore not meaningful until the
estimated profiles have been aligned with their true counterparts.

Let \(\mathcal{P}_K\) denote the set of all permutations of
\(\{1,\ldots,K\}\). The alignment permutation is selected by minimizing the
total squared distance between the estimated and true component means:
\begin{equation}
\widehat{\sigma}
=
\underset{\sigma\in\mathcal{P}_K}{\operatorname{argmin}}
\sum_{k=1}^{K}
\left\|
\widehat{\mu}_{\sigma(k)}
-
\mu_k^\ast
\right\|^2.
\label{eq:recovery_label_alignment}
\end{equation}

The symbol \(\sigma\) is used for the permutation in order to distinguish the
label-alignment rule from the joint posterior responsibilities
\(\tau_{ik}\).

After determining \(\widehat{\sigma}\), all estimated component-specific
quantities are reordered consistently:
\begin{equation}
\left(
\widehat{\pi}_{\sigma(k)},
\widehat{\mu}_{\sigma(k)},
\widehat{\Sigma}_{\sigma(k)},
\widehat{p}_{\sigma(k)}
\right)
\longrightarrow
\left(
\widehat{\pi}_k,
\widehat{\mu}_k,
\widehat{\Sigma}_k,
\widehat{p}_k
\right).
\label{eq:recovery_parameter_relabeling}
\end{equation}

The columns of the joint responsibilities and score-based posterior matrices
are reordered using the same permutation:
\begin{equation}
\widehat{\tau}_{i,\sigma(k)}
\longrightarrow
\widehat{\tau}_{ik},
\qquad
\widehat{\gamma}_{i,\sigma(k)}^{X}
\longrightarrow
\widehat{\gamma}_{ik}^{X}.
\label{eq:recovery_posterior_relabeling}
\end{equation}

All component-specific comparisons reported in
Section~\ref{sec:simulation_results} are performed after label alignment.

\subsection{Validation targets}
\label{subsec:recovery_validation_targets}

The recovery experiment evaluates four distinct levels of agreement between
the estimated model and the data-generating process.

\subsubsection{Latent mixture structure}

The first level concerns the recovery of
\begin{equation}
\left\{
\pi_k^\ast,
\mu_k^\ast,
\Sigma_k^\ast
\right\}_{k=1}^{K}
\label{eq:recovery_target_mixture}
\end{equation}
through the corresponding estimates
\begin{equation}
\left\{
\widehat{\pi}_k,
\widehat{\mu}_k,
\widehat{\Sigma}_k
\right\}_{k=1}^{K}.
\label{eq:recovery_estimated_mixture}
\end{equation}

This comparison evaluates whether the joint EM procedure reconstructs the
latent population structure underlying the observed score distribution.

\subsubsection{Profile-specific probabilities of default}

The second level concerns the recovery of
\begin{equation}
p_k^\ast
=
P_{\Theta^\ast}(D_i=1\mid Z_i=k)
\label{eq:recovery_target_profile_pd}
\end{equation}
through
\begin{equation}
\widehat p_k
=
P_{\widehat\Theta}(D_i=1\mid Z_i=k).
\label{eq:recovery_estimated_profile_pd_target}
\end{equation}

This comparison evaluates whether the realized default outcomes and estimated
latent responsibilities are sufficient to reconstruct the structural default
risk associated with each latent profile.

\subsubsection{Score-based posterior beliefs}

The third level concerns the borrower-level comparison
\begin{equation}
\gamma_i^{X,\ast}
\quad\text{versus}\quad
\widehat{\gamma}_i^{X}.
\label{eq:recovery_target_posterior}
\end{equation}

This is the central validation target of the paper. It evaluates whether the
estimation procedure recovers the complete distribution representing the
intermediary's uncertainty about the latent economic state of the borrower.

The comparison is performed between posterior distributions, rather than only
between the corresponding modal assignments
\begin{equation}
Z_i^\ast
\quad\text{and}\quad
\widehat Z_i
=
\underset{k}{\operatorname{argmax}}\,
\widehat{\gamma}_{ik}^{X}.
\label{eq:recovery_modal_classification}
\end{equation}

Hard classification accuracy may provide a useful complementary diagnostic,
but it does not preserve the uncertainty contained in the full posterior
vector.

\subsubsection{Borrower-level probabilities of default}

The fourth level concerns the comparison
\begin{equation}
PD_i^\ast
\quad\text{versus}\quad
\widehat{PD}_i.
\label{eq:recovery_target_borrower_pd}
\end{equation}

This comparison evaluates whether the joint recovery of
\(\widehat p_k\) and \(\widehat\gamma_{ik}^{X}\) translates into an accurate
borrower-level credit-risk measure.

\subsection{Information-geometric validation strategy}
\label{subsec:recovery_information_geometry}

For every borrower, the true and estimated score-based posterior vectors belong
to the probability simplex
\begin{equation}
\Delta^{K-1}
=
\left\{
q\in\mathbb{R}^{K}:
q_k\geq 0,\,
\sum_{k=1}^{K}q_k=1
\right\}.
\label{eq:recovery_probability_simplex}
\end{equation}

The posterior recovery exercise can therefore be interpreted geometrically as
a comparison between two points on the same statistical space:
\begin{equation}
\gamma_i^{X,\ast},
\widehat{\gamma}_i^{X}
\in
\Delta^{K-1}.
\label{eq:recovery_posterior_simplex_points}
\end{equation}

The validation reported in Section~\ref{sec:simulation_results} combines
Euclidean recovery measures with information-geometric divergences. In
particular, posterior recovery is evaluated through:

\begin{itemize}
    \item the mean squared distance between true and estimated posterior
    vectors;
    \item the average Kullback--Leibler divergence;
    \item the average Jensen--Shannon divergence.
\end{itemize}

The Kullback--Leibler divergence preserves the directional interpretation of
informational loss, while the Jensen--Shannon divergence provides a symmetric
and numerically bounded comparison between posterior distributions.

The same posterior beliefs are also used to evaluate the information loss
induced by the product-of-marginals approximation, in which the joint posterior over
latent creditworthiness and financial fragility is replaced by the product of
its marginal distributions.

\subsection{Summary of the recovery design}
\label{subsec:recovery_design_summary}

The recovery experiment follows the sequence
\begin{equation}
\{(X_i,D_i)\}_{i=1}^{n}
\longrightarrow
\widehat{\Theta}
\longrightarrow
\widehat{\gamma}_i^{X}
\longrightarrow
\widehat{PD}_i.
\label{eq:recovery_workflow}
\end{equation}

The first step uses historical scores and realized default outcomes to estimate
the latent mixture structure and profile-specific default probabilities. The
second step reconstructs the predictive posterior distribution using only the
observed score vector. The third step translates this posterior distribution
into a borrower-level probability of default.

The validation then compares the estimated objects with their known
data-generating counterparts:
\begin{equation}
\widehat{\Theta}
\quad\text{versus}\quad
\Theta^\ast,
\label{eq:recovery_parameter_comparison}
\end{equation}
\begin{equation}
\widehat{\gamma}_i^{X}
\quad\text{versus}\quad
\gamma_i^{X,\ast},
\label{eq:recovery_posterior_comparison}
\end{equation}
and
\begin{equation}
\widehat{PD}_i
\quad\text{versus}\quad
PD_i^\ast.
\label{eq:recovery_pd_comparison}
\end{equation}

This structure separates three conceptually distinct questions: whether the
structural parameters of the latent model can be recovered, whether the
borrower-level information states can be reconstructed, and whether those
recovered information states produce accurate probabilities of default.

The numerical results of these comparisons are reported in
Section~\ref{sec:simulation_results}.

\section{Simulation Results}
\label{sec:simulation_results}

This section reports the results of the controlled simulation experiment specified in Section~\ref{sec:first_simulation}, estimated using the joint EM procedure developed in Section~\ref{sec:em}, and evaluated according to the recovery design described in Section~\ref{sec:em_recovery}.

The purpose of the exercise is to
assess whether the joint Expectation--Maximization procedure is able to recover
the latent population structure, the profile-specific probabilities of default,
and the posterior distribution representing the intermediary's beliefs about
the latent state of each borrower.

The experiment distinguishes between two posterior distributions. During
estimation, the EM algorithm computes the responsibilities
\[
\tau_{ik}
=
P(Z_i=k\mid X_i,D_i),
\]
which use the observed score vector and the realized default outcome. After
estimation, borrower-level posterior beliefs are computed as
\[
\widehat{\gamma}_{ik}^{X}
=
P_{\widehat{\Theta}}(Z_i=k\mid X_i),
\]
using only the observed score vector. The latter distribution represents the
information available for predictive credit-risk assessment and is therefore
used to calculate borrower-level probabilities of default.

All numerical results reported below are obtained from the baseline simulation
design, using \(n=5{,}000\) borrowers and \(K=4\) latent profiles. The true
parameter values are known by construction and provide the benchmark against
which the estimated quantities are evaluated.

All recovery metrics reported in this section are computed from the unrounded parameter estimates.

\subsection{Recovery of the latent mixture structure}
\label{subsec:results_mixture_recovery}

The first validation exercise concerns the recovery of the Gaussian mixture
parameters. The true data-generating parameter set for the observed score
distribution is
\[
\Theta_X^\ast
=
\left\{
\pi_k^\ast,\mu_k^\ast,\Sigma_k^\ast
\right\}_{k=1}^{K}.
\]

The joint EM procedure produces the corresponding estimates
\[
\widehat{\Theta}_X
=
\left\{
\widehat{\pi}_k,\widehat{\mu}_k,\widehat{\Sigma}_k
\right\}_{k=1}^{K}.
\]

Since the likelihood of a finite mixture is invariant with respect to
permutations of the component labels, the estimated components are aligned with
the true latent profiles before performing component-specific comparisons.

\subsubsection{Mixture probabilities}

Table~\ref{tab:results_pi} compares the true and estimated mixture probabilities.

\begin{table}[htbp]
\centering
\caption{Recovery of the latent mixture probabilities}
\label{tab:results_pi}
\begin{tabular}{cccc}
\hline
Latent profile \(k\)
&
True \(\pi_k^\ast\)
&
Estimated \(\widehat{\pi}_k\)
&
Absolute error
\\
\hline
1 & 0.400 & 0.391 & 0.009 \\
2 & 0.200 & 0.200 & 0.000 \\
3 & 0.250 & 0.254 & 0.004 \\
4 & 0.150 & 0.155 & 0.005\\
\hline
\end{tabular}
\end{table}

The overall recovery error for the mixture probabilities is summarized by
\begin{equation}
\operatorname{MAE}_{\pi}
=
\frac{1}{K}
\sum_{k=1}^{K}
\left|
\widehat{\pi}_k-\pi_k^\ast
\right|.
\label{eq:results_mae_pi}
\end{equation}

For the baseline experiment, the estimated value is
\[
\operatorname{MAE}_{\pi}=4.591\times10^{-3}.
\]

A small value of \(\operatorname{MAE}_{\pi}\) indicates that the joint EM
procedure correctly recovers the relative prevalence of the latent borrower
profiles in the simulated population.

\subsubsection{Component-specific means}

Table~\ref{tab:results_mu} compares the true and estimated component-specific
mean vectors.

\begin{table}[htbp]
\centering
\caption{Recovery of the component-specific mean vectors}
\label{tab:results_mu}
\begin{tabular}{ccccc}
\hline
Profile
&
\(\mu_{R,k}^\ast\)
&
\(\mu_{F,k}^\ast\)
&
\(\widehat{\mu}_{R,k}\)
&
\(\widehat{\mu}_{F,k}\)
\\
\hline
1 & 1.800 & 0.300 & 1.792 & 0.280\\
2 & 1.800 & 1.800 & 1.849 & 1.808\\
3 & -0.800 & 0.300 & -0.773 & 0.298 \\
4 & -0.800 & 1.800 & -0.811 & 1.771 \\
\hline
\end{tabular}
\end{table}

The average squared distance between estimated and true component means is
\begin{equation}
\operatorname{MSE}_{\mu}
=
\frac{1}{K}
\sum_{k=1}^{K}
\left\|
\widehat{\mu}_k-\mu_k^\ast
\right\|^2.
\label{eq:results_mse_mu}
\end{equation}

For the baseline experiment,
\[
\operatorname{MSE}_{\mu}=1.170\times10{-3}.
\]

\subsubsection{Covariance matrices}

The recovery of the component-specific covariance matrices is evaluated using
the squared Frobenius norm:
\begin{equation}
\operatorname{MSE}_{\Sigma}
=
\frac{1}{K}
\sum_{k=1}^{K}
\left\|
\widehat{\Sigma}_k-\Sigma_k^\ast
\right\|_F^2.
\label{eq:results_mse_sigma}
\end{equation}

The resulting value is
\[
\operatorname{MSE}_{\Sigma}=2.157\times 10{-3}.
\]

Taken together, the recovery of
\(\pi_k^\ast\), \(\mu_k^\ast\), and \(\Sigma_k^\ast\) indicates whether the
estimated model reproduces the main characteristics of the latent population
generating the observed score distribution. Parameter recovery, however,
represents only the first level of validation. The extended estimation
procedure must also recover the structural relationship between latent borrower
profiles and default.

\subsection{Recovery of profile-specific probabilities of default}
\label{subsec:results_profile_pd}

The second validation exercise concerns the profile-specific probabilities of
default. In the data-generating process, these parameters are defined as
\begin{equation}
p_k^\ast
=
P_{\Theta^\ast}(D_i=1\mid Z_i=k),
\qquad
k=1,\ldots,K.
\label{eq:results_true_profile_pd}
\end{equation}

The baseline parameter vector is
\begin{equation}
\mathbf{p}^\ast
=
\left(
0.01,\,
0.05,\,
0.10,\,
0.25
\right)^\top.
\label{eq:results_true_profile_pd_vector}
\end{equation}

Unlike the data-generating process, the estimation procedure treats these
probabilities as unknown. The joint EM algorithm estimates them using the
responsibility-weighted default rates:
\begin{equation}
\widehat{p}_k
=
\frac{
\displaystyle
\sum_{i=1}^{n}
\widehat{\tau}_{ik}D_i
}{
\displaystyle
\sum_{i=1}^{n}
\widehat{\tau}_{ik}
}.
\label{eq:results_estimated_profile_pd}
\end{equation}

Table~\ref{tab:results_profile_pd} compares the true and estimated
profile-specific probabilities of default.

\begin{table}[htbp]
\centering
\caption{Recovery of the profile-specific probabilities of default}
\label{tab:results_profile_pd}
\begin{tabular}{cccc}
\hline
Latent profile \(k\)
&
True \(p_k^\ast\)
&
Estimated \(\widehat{p}_k\)
&
Absolute error
\\
\hline
1 & 0.010 & 0.011 & 0.001\\
2 & 0.050 & 0.051 & 0.001 \\
3 & 0.100 & 0.104 & 0.004 \\
4 & 0.250 & 0.242 & 0.008 \\
\hline
\end{tabular}
\end{table}

The overall recovery of the structural default parameters is assessed through
\begin{equation}
\operatorname{MAE}_{p}
=
\frac{1}{K}
\sum_{k=1}^{K}
\left|
\widehat{p}_k-p_k^\ast
\right|
\label{eq:results_profile_pd_mae}
\end{equation}
and
\begin{equation}
\operatorname{MSE}_{p}
=
\frac{1}{K}
\sum_{k=1}^{K}
\left(
\widehat{p}_k-p_k^\ast
\right)^2.
\label{eq:results_profile_pd_mse}
\end{equation}

For the baseline simulation, the resulting values are
\[
\operatorname{MAE}_{p}=3.263\times10{-3} 
\]
and
\[
\operatorname{MSE}_{p}=1.902\times 10^{-5}.
\]

The recovery of \(p_k^\ast\) is an important part of the validation exercise.
It shows whether the realized default outcomes, combined with the estimated
latent responsibilities, are sufficient to recover the structural default rate
associated with each latent borrower profile.

Estimation accuracy may differ across profiles because the effective number of
observations and the number of realized defaults are not the same for all
components. In particular, the estimation of low probabilities of default may
be less precise because the corresponding latent profiles generate relatively
few default events.

\subsection{Recovery of posterior beliefs}
\label{subsec:results_posterior_recovery}

The third validation exercise evaluates whether the estimated model recovers the
posterior distribution of the latent states based on the observed score vector.

For each borrower, the true posterior belief is
\begin{equation}
\gamma_{ik}^{X,\ast}
=
P_{\Theta^\ast}(Z_i=k\mid X_i),
\label{eq:results_true_score_posterior}
\end{equation}
while the estimated posterior belief is
\begin{equation}
\widehat{\gamma}_{ik}^{X}
=
P_{\widehat{\Theta}}(Z_i=k\mid X_i).
\label{eq:results_estimated_score_posterior}
\end{equation}

The comparison is performed at the level of the complete posterior vectors
\[
\gamma_i^{X,\ast}
=
\left(
\gamma_{i1}^{X,\ast},
\ldots,
\gamma_{iK}^{X,\ast}
\right)
\]
and
\[
\widehat{\gamma}_i^{X}
=
\left(
\widehat{\gamma}_{i1}^{X},
\ldots,
\widehat{\gamma}_{iK}^{X}
\right).
\]

It is important that this comparison uses score-based posterior beliefs. The
realized default outcome \(D_i\) is used during training to estimate the model
parameters, but it is not used to construct the predictive posterior belief
\(\widehat{\gamma}_i^{X}\).

The average squared posterior recovery error is
\begin{equation}
\operatorname{MSE}_{\gamma}
=
\frac{1}{n}
\sum_{i=1}^{n}
\left\|
\widehat{\gamma}_i^{X}
-
\gamma_i^{X,\ast}
\right\|^2.
\label{eq:results_posterior_mse}
\end{equation}

For the baseline simulation,

\[ \operatorname{MSE}_{\gamma} = 8.163 \times 10^{-4}. \]

\subsection{Information-geometric validation}
\label{subsec:results_information_geometry}

Both the true and estimated posterior vectors belong to the probability simplex
\[
\Delta^{K-1}
=
\left\{
q\in\mathbb{R}^{K}:
q_k\geq 0,\,
\sum_{k=1}^{K}q_k=1
\right\}.
\]

The discrepancy between the true and estimated posterior beliefs can therefore
be assessed through information-geometric quantities.

For borrower \(i\), the Kullback--Leibler divergence is
\begin{equation}
D_{\mathrm{KL}}
\left(
\gamma_i^{X,\ast}
\,\middle\|\,
\widehat{\gamma}_i^{X}
\right)
=
\sum_{k=1}^{K}
\gamma_{ik}^{X,\ast}
\log
\left(
\frac{
\gamma_{ik}^{X,\ast}
}{
\widehat{\gamma}_{ik}^{X}
}
\right).
\label{eq:results_individual_kl}
\end{equation}

The average Kullback--Leibler divergence is
\begin{equation}
\overline{D}_{\mathrm{KL}}
=
\frac{1}{n}
\sum_{i=1}^{n}
D_{\mathrm{KL}}
\left(
\gamma_i^{X,\ast}
\,\middle\|\,
\widehat{\gamma}_i^{X}
\right).
\label{eq:results_average_kl}
\end{equation}

Since the Kullback--Leibler divergence is asymmetric and can be sensitive to
probabilities close to zero, we also calculate the Jensen--Shannon divergence.
Define
\begin{equation}
m_i
=
\frac{1}{2}
\left(
\gamma_i^{X,\ast}
+
\widehat{\gamma}_i^{X}
\right).
\label{eq:results_js_midpoint}
\end{equation}

The borrower-level Jensen--Shannon divergence is
\begin{align}
D_{\mathrm{JS}}
\left(
\gamma_i^{X,\ast},
\widehat{\gamma}_i^{X}
\right)
={}&
\frac{1}{2}
D_{\mathrm{KL}}
\left(
\gamma_i^{X,\ast}
\,\middle\|\,
m_i
\right)
\nonumber\\
&+
\frac{1}{2}
D_{\mathrm{KL}}
\left(
\widehat{\gamma}_i^{X}
\,\middle\|\,
m_i
\right).
\label{eq:results_individual_js}
\end{align}

The corresponding average is
\begin{equation}
\overline{D}_{\mathrm{JS}}
=
\frac{1}{n}
\sum_{i=1}^{n}
D_{\mathrm{JS}}
\left(
\gamma_i^{X,\ast},
\widehat{\gamma}_i^{X}
\right).
\label{eq:results_average_js}
\end{equation}

Table~\ref{tab:results_posterior} reports the posterior recovery metrics.

\begin{table}[htbp]
\centering
\caption{Recovery of borrower-level posterior beliefs}
\label{tab:results_posterior}
\begin{tabular}{cc}
\hline
Measure & Estimated value \\
\hline
\(\operatorname{MSE}_{\gamma}\) & $8.163\times10{-4}$ \\
\(\overline{D}_{\mathrm{KL}}\) & $1.736\times10{-3}$ \\
\(\overline{D}_{\mathrm{JS}}\) & $4.367\times10{-4}$\\
\hline
\end{tabular}
\end{table}

Small values of these measures indicate that the estimated posterior
distributions are close to the true information states generated by the model.
This comparison is more informative than hard classification accuracy because
it evaluates whether the estimation procedure reproduces the full distribution
of uncertainty over the latent borrower profiles.

\subsection{Recovery of borrower-level probabilities of default}
\label{subsec:results_borrower_pd}

The fourth validation exercise examines whether the joint recovery of the
profile-specific default probabilities and score-based posterior beliefs
translates into accurate borrower-level probabilities of default.

Under the true data-generating parameters, the borrower-level probability of
default is
\begin{equation}
PD_i^\ast
=
\sum_{k=1}^{K}
p_k^\ast
\gamma_{ik}^{X,\ast}.
\label{eq:results_true_borrower_pd}
\end{equation}

The corresponding estimated probability is
\begin{equation}
\widehat{PD}_i
=
\sum_{k=1}^{K}
\widehat{p}_k
\widehat{\gamma}_{ik}^{X}.
\label{eq:results_estimated_borrower_pd}
\end{equation}

Unlike an evaluation in which the true \(p_k^\ast\) are retained in the
estimated PD formula, Equation~\eqref{eq:results_estimated_borrower_pd} uses
only estimated quantities. Consequently, the difference between
\(\widehat{PD}_i\) and \(PD_i^\ast\) reflects both posterior recovery error and
profile-specific default-probability estimation error.

The borrower-level prediction errors are summarized by
\begin{equation}
\operatorname{MSE}_{PD}
=
\frac{1}{n}
\sum_{i=1}^{n}
\left(
\widehat{PD}_i-PD_i^\ast
\right)^2
\label{eq:results_borrower_pd_mse}
\end{equation}
and
\begin{equation}
\operatorname{MAE}_{PD}
=
\frac{1}{n}
\sum_{i=1}^{n}
\left|
\widehat{PD}_i-PD_i^\ast
\right|.
\label{eq:results_borrower_pd_mae}
\end{equation}

The resulting values are reported in
Table~\ref{tab:borrower_pd_recovery}.

\begin{table}[htbp] 
\centering 
\caption{Recovery of borrower-level probabilities of default} \label{tab:borrower_pd_recovery} 
\begin{tabular}{lc} 
\toprule Measure & Estimated value \\ 
\midrule $\operatorname{MSE}_{\mathrm{PD}}$ & $1.419 \times 10^{-5}$ \\ $\operatorname{MAE}_{\mathrm{PD}}$ & $2.691 \times 10^{-3}$ \\ 
\bottomrule 
\end{tabular} 
\end{table}

The correspondence between \(PD_i^\ast\) and \(\widehat{PD}_i\) provides a
synthetic validation of the complete estimation procedure. Accurate recovery
requires both the correct estimation of the structural default probabilities
and the correct reconstruction of the borrower-level posterior beliefs.

The result also illustrates the economic role of the posterior distribution.
Although the structural default probability is constant within each latent
profile, the borrower-level probability of default varies across observations
because different score vectors imply different posterior distributions over
the latent states.

\subsection{Decomposition of borrower-level PD error}
\label{subsec:results_pd_decomposition}

To distinguish the two sources of borrower-level PD error, consider the
intermediate quantity
\begin{equation}
\widetilde{PD}_i
=
\sum_{k=1}^{K}
p_k^\ast
\widehat{\gamma}_{ik}^{X}.
\label{eq:results_intermediate_pd}
\end{equation}

This quantity uses the estimated posterior beliefs but retains the true
profile-specific default probabilities. It isolates the effect of posterior
recovery error.

The overall difference can be decomposed as
\begin{align}
\widehat{PD}_i-PD_i^\ast
={}&
\left(
\widehat{PD}_i-\widetilde{PD}_i
\right)
+
\left(
\widetilde{PD}_i-PD_i^\ast
\right)
\nonumber\\
={}&
\sum_{k=1}^{K}
\left(
\widehat{p}_k-p_k^\ast
\right)
\widehat{\gamma}_{ik}^{X}
+
\sum_{k=1}^{K}
p_k^\ast
\left(
\widehat{\gamma}_{ik}^{X}
-
\gamma_{ik}^{X,\ast}
\right).
\label{eq:results_pd_error_decomposition}
\end{align}

The first term captures the contribution of profile-specific PD estimation
error, while the second term captures the contribution of posterior belief
recovery error.

For descriptive purposes, the two components can be summarized through
\begin{equation}
\operatorname{MAE}_{PD}^{(p)}
=
\frac{1}{n}
\sum_{i=1}^{n}
\left|
\widehat{PD}_i-\widetilde{PD}_i
\right|
\label{eq:results_pd_error_profile_component}
\end{equation}
and
\begin{equation}
\operatorname{MAE}_{PD}^{(\gamma)}
=
\frac{1}{n}
\sum_{i=1}^{n}
\left|
\widetilde{PD}_i-PD_i^\ast
\right|.
\label{eq:results_pd_error_posterior_component}
\end{equation}

Table~\ref{tab:results_pd_decomposition} reports these descriptive measures.

\begin{table}[htbp]
\centering
\caption{Decomposition of borrower-level PD recovery error}
\label{tab:results_pd_decomposition}
\begin{tabular}{lc}
\hline
Error component & Estimated value \\
\hline
Overall \(\operatorname{MAE}_{PD}\) & $2.691\times 10^{-3}$ \\
Profile-PD component \(\operatorname{MAE}_{PD}^{(p)}\) & $2.005\times 10^{-3}$ \\
Posterior component \(\operatorname{MAE}_{PD}^{(\gamma)}\) & $1.082\times 10^{-3}$ \\
\hline
\end{tabular}
\end{table}

Because absolute values prevent a strictly additive decomposition of the average
absolute error, the last two measures should be interpreted as descriptive
indicators of the magnitude of the two error channels rather than as components
that necessarily sum to the total \(\operatorname{MAE}_{PD}\).

\subsection{Product-of-marginals approximation and posterior dependence}
\label{subsec:results_mean_field}

The final validation exercise evaluates the information loss associated with
replacing the joint posterior over latent creditworthiness and financial
fragility with a product-of-marginals representation.

Let 
\begin{equation} 
\widehat{p}_i^{X}(z_R,z_F) = P_{\widehat{\boldsymbol{\Theta}}} \left( Z_R=z_R, Z_F=z_F \mid \boldsymbol{X}_i \right),
\label{eq:estimated_joint_score_posterior} \end{equation} 
denote the estimated joint score-based posterior. 
The corresponding estimated marginal posterior distributions are \begin{equation} \widehat{p}_{R,i}^{X}(z_R) = \sum_{z_F} \widehat{p}_i^{X}(z_R,z_F) \label{eq:estimated_creditworthiness_marginal} \end{equation} and \begin{equation} \widehat{p}_{F,i}^{X}(z_F) = \sum_{z_R} \widehat{p}_i^{X}(z_R,z_F). \label{eq:estimated_fragility_marginal} \end{equation} The estimated product-of-marginals approximation is then \begin{equation} \widehat{q}_i^{\mathrm{PM}}(z_R,z_F) = \widehat{p}_{R,i}^{X}(z_R) \widehat{p}_{F,i}^{X}(z_F). \label{eq:estimated_product_of_marginals} \end{equation}

The informational loss induced by the factorization is measured by

\begin{equation} \begin{aligned} D_{\mathrm{KL}} \left( \widehat{p}_i^{X} \middle\| \widehat{q}_i^{\mathrm{PM}} \right) = \sum_{z_R} \sum_{z_F} \widehat{p}_i^{X}(z_R,z_F) \log \left[ \frac{ \widehat{p}_i^{X}(z_R,z_F) }{ \widehat{p}_{R,i}^{X}(z_R) \widehat{p}_{F,i}^{X}(z_F) } \right]. \end{aligned} 
\label{eq:results_mean_field_kl}
\end{equation}

The average product-of-marginals information loss is
\begin{equation} \overline{D}_{\mathrm{KL}}^{\mathrm{PM}} = \frac{1}{n} \sum_{i=1}^{n} D_{\mathrm{KL}} \left( \widehat{p}_i^{X} \middle\| \widehat{q}_i^{\mathrm{PM}} \right). \label{eq:average_product_marginals_kl} \end{equation}

The borrower-level probability of default implied by the product-of-marginals
approximation is
\begin{equation} \widehat{\mathrm{PD}}_i^{\mathrm{PM}} = \sum_{z_R} \sum_{z_F} \widehat{q}_i^{\mathrm{PM}}(z_R,z_F) \widehat{p}_{z_R,z_F}, \end{equation}

where \(\widehat{p}_{z_R,z_F}\) denotes the estimated default probability
associated with the corresponding joint latent profile.

The effect of the approximation on borrower-level PD is measured by
\begin{equation}
\operatorname{MAE}_{PD}^{PM}
=
\frac{1}{n}
\sum_{i=1}^{n}
\left|
\widehat{PD}_i^{PM}
-
\widehat{PD}_i
\right|.
\label{eq:results_mean_field_pd_mae}
\end{equation}

Table~\ref{tab:results_mean_field} summarizes the product-of-marginals results.

\begin{table}[htbp]
\centering
\caption{Information loss under the product-of-marginals approximation}
\label{tab:results_mean_field}
\begin{tabular}{cc}
\hline
Measure & Estimated value \\
\hline
\(\overline{D}_{\mathrm{KL}}^{PM}\) & $2.474\times 10^{-3}$ \\
\(\operatorname{MAE}_{PD}^{PM}\) & $2.193\times 10^{-4}$ \\
\hline
\end{tabular}
\end{table}

These measures quantify the information discarded when the joint posterior is
replaced by separate beliefs about creditworthiness and financial fragility. 

At borrower level, the divergence between the joint posterior and the product of its marginal distributions is equal to the posterior mutual information between latent creditworthiness and financial fragility. Its sample average therefore measures the average residual dependence between the two latent dimensions under the estimated score-based posterior.

\subsection{Summary of the simulation evidence}
\label{subsec:results_summary}

Table~\ref{tab:results_summary} 
summarizes the main validation results.

\begin{table}[htbp] 
\centering 
\caption{Summary of the simulation results} \label{tab:results_summary} 
\begin{tabular}{lc} 
\toprule Validation target & Estimated value \\ 
\midrule 
$\operatorname{MAE}_{\pi}$ & $4.591 \times 10^{-3}$ \\ $\operatorname{MSE}_{\mu}$ & $1.170 \times 10^{-3}$ \\ $\operatorname{MSE}_{\Sigma}$ & $2.157 \times 10^{-3}$ \\ $\operatorname{MAE}_{p}$ & $3.263 \times 10^{-3}$ \\ $\operatorname{MSE}_{p}$ & $1.902 \times 10^{-5}$ \\ $\operatorname{MSE}_{\gamma}$ & $8.163 \times 10^{-4}$ \\ $\overline{D}_{\mathrm{KL}}$ & $1.736 \times 10^{-3}$ \\ $\overline{D}_{\mathrm{JS}}$ & $4.367 \times 10^{-4}$ \\ $\operatorname{MSE}_{\mathrm{PD}}$ & $1.419 \times 10^{-5}$ \\ $\operatorname{MAE}_{\mathrm{PD}}$ & $2.691 \times 10^{-3}$ \\ $\overline{D}_{\mathrm{KL}}^{\mathrm{PM}}$ & $2.474 \times 10^{-3}$ \\ $\operatorname{MAE}_{\mathrm{PD}}^{\mathrm{PM}}$ & $2.193 \times 10^{-4}$ 
\\ \bottomrule 
\end{tabular} 
\end{table}

The simulation evidence provides four distinct levels of validation. First, the comparison of the Gaussian mixture parameters evaluates whether the EM procedure recovers the latent population structure underlying the observed score distribution. Second, the comparison of $p_k^{*}$ and $\widehat{p}_k$ evaluates whether the realized default outcomes are sufficient to recover the structural default probabilities associated with the latent profiles. Third, the comparison of $\boldsymbol{\gamma}_i^{X,*}$ and $\widehat{\boldsymbol{\gamma}}_i^{X}$ evaluates whether the estimated model reconstructs the complete borrower-level information state. Fourth, the comparison of $\mathrm{PD}_i^{*}$ and $\widehat{\mathrm{PD}}_i$ evaluates whether posterior and structural parameter recovery translate into accurate borrower-level probabilities of default.

The distinction between the joint estimation responsibilities \[ \widehat{\tau}_{ik} = P_{\widehat{\Theta}}(Z_i=k\mid X_i,D_i) \] and the score-based posterior beliefs \[ \widehat{\gamma}_{ik}^{X} = P_{\widehat{\Theta}}(Z_i=k\mid X_i) \] is central to the interpretation of the exercise.

The first distribution is an
internal object used to estimate the model from historical observations. The
second distribution represents the information available before the realization
of default and is therefore the relevant posterior for borrower-level credit-risk
prediction.

Overall, successful recovery requires more than identifying the most likely
latent profile. It requires recovering the structural default probabilities and
the full posterior distribution over latent states. The estimated borrower-level
probability of default is then obtained as a functional of these two recovered
objects:
\begin{equation}
\widehat{PD}_i
=
\sum_{k=1}^{K}
\widehat{p}_k
\widehat{\gamma}_{ik}^{X}.
\label{eq:results_final_pd_representation}
\end{equation}

This result supports the central interpretation proposed in the paper. Credit
risk is not represented only by a deterministic assignment to a latent profile
or by a direct mapping from observed scores to default. It is obtained by
combining the estimated structural risk associated with each latent state with
the posterior belief distribution representing the intermediary's uncertainty
about the borrower's latent economic condition.

The simulation should nevertheless be interpreted as a controlled recovery
experiment. Because the true parameters and posterior distributions are known
by construction, it provides a benchmark for evaluating the internal
consistency of the proposed estimation framework. The experiment does not, by
itself, establish performance on real credit portfolios. Empirical calibration
and out-of-sample validation remain necessary extensions of the present analysis.

The results establish the internal coherence of the proposed recovery framework under the controlled data-generating conditions considered in this experiment. Their broader interpretation, the limitations of the simulation design, and the implications for empirical credit-risk modelling are discussed in the following section.

\section{Discussion} \label{sec:discussion} The simulation results support the internal coherence of the proposed latent-state framework. Under the controlled data-generating conditions considered in this paper, the joint EM procedure accurately recovers the Gaussian mixture structure, the profile-specific probabilities of default, and the score-based posterior distributions over the latent borrower profiles. The limited errors observed in the resulting borrower-level probabilities of default indicate that the two main estimated objects, \[ \widehat{\boldsymbol{p}} \qquad\text{and}\qquad \widehat{\boldsymbol{\gamma}}^{X}_{i}, \] are jointly reconstructed with satisfactory accuracy. The central implication of the experiment is that posterior belief recovery and latent-profile classification are distinct statistical objectives. A hard classification assigns each borrower to the most likely latent profile but discards the uncertainty associated with the remaining profiles. By contrast, the complete posterior vector \[ \widehat{\boldsymbol{\gamma}}^{X}_{i} = \left( \widehat{\gamma}^{X}_{i1}, \ldots, \widehat{\gamma}^{X}_{iK} \right) \] preserves the intermediary's uncertainty concerning the latent economic condition of the borrower. This information is relevant because borrowers assigned to the same modal profile may nevertheless have different posterior distributions and, consequently, different borrower-level probabilities of default. An important feature of the framework is the distinction between the posterior responsibilities used during estimation and the posterior beliefs used for predictive credit-risk assessment. The EM responsibilities \[ \widehat{\tau}_{ik} = P_{\widehat{\Theta}} \left( Z_i=k\mid X_i,D_i \right) \] use both historical scores and realized default outcomes to estimate the model parameters. After estimation, the relevant borrower-level information state is instead \[ \widehat{\gamma}^{X}_{ik} = P_{\widehat{\Theta}} \left( Z_i=k\mid X_i \right), \] which is conditional only on the information available before default realization. The realized outcome therefore contributes to historical parameter estimation without being used as a predictor of its own borrower-level probability of default. The decomposition of borrower-level PD error also clarifies the interpretation of the results. The difference between estimated and true PDs reflects both the error in recovering the posterior beliefs and the error in estimating the profile-specific default probabilities: \[ \widehat{\operatorname{PD}}_i-\operatorname{PD}^{*}_i = \sum_{k=1}^{K} \left( \widehat{p}_k-p^{*}_k \right) \widehat{\gamma}^{X}_{ik} + \sum_{k=1}^{K} p^{*}_k \left( \widehat{\gamma}^{X}_{ik} - \gamma^{X,*}_{ik} \right). \] The limited magnitude of both components in the baseline experiment shows that borrower-level PD recovery is not produced by a compensation between large estimation errors in the two underlying objects. The product-of-marginals results provide a complementary insight. Replacing the joint posterior distribution by the product of its marginal distributions generates a positive information loss, although its effect on borrower-level PD is limited in the baseline experiment. This indicates that an approximation may alter the posterior distribution without producing an equally large change in a particular functional of that distribution. The quality of a posterior approximation should therefore be evaluated both through distributional divergences and through its effect on the credit-risk quantity of interest. These findings must nevertheless be interpreted within the limits of the simulation design. The data are generated from the same parametric family used for estimation, the number of latent profiles is known, and the Gaussian components are sufficiently separated to permit reliable identification. Moreover, the structural assumption \[ D_i \indep X_i \mid Z_i \] holds by construction. The experiment consequently validates the internal recovery properties of the procedure but does not establish its empirical performance on real credit portfolios. In an empirical application, the number and interpretation of the latent profiles would not be known in advance, score distributions could depart from Gaussianity, and default could retain a direct dependence on observed borrower characteristics after conditioning on the latent state. The stability of the estimated profiles, the calibration of borrower-level probabilities of default, and the robustness of posterior beliefs would therefore require explicit out-of-sample and temporal validation. These issues motivate the subsequent application of the framework to real borrower-level credit data.

\section{Conclusions} \label{sec:conclusions} This paper has developed a latent-state framework for recovering borrower-level posterior beliefs in credit-risk analysis. The proposed joint EM procedure estimates the Gaussian mixture structure and the profile-specific probabilities of default from observed borrower-level scores and realized default outcomes. After estimation, the posterior beliefs relevant for credit-risk assessment are computed conditionally on the observed score vector alone. The controlled simulation shows that the proposed procedure can recover the latent mixture parameters, the profile-specific probabilities of default, the borrower-level posterior beliefs, and the resulting probabilities of default with limited estimation error. The results support the central interpretation of the paper: borrower-level credit risk can be represented through a posterior distribution over latent economic states rather than only through a hard classification or a point-valued score. The present analysis remains a controlled recovery experiment in which the data-generating process and the number of latent profiles are known by construction. A subsequent paper will extend the framework to real borrower-level credit data, with the objective of assessing its empirical estimation, stability, calibration, and predictive performance.

\clearpage 
\printbibliography

\appendix

\section{Reproducibility and Ancillary Code}

All results are based on synthetic data generated with a fixed random seed. The accompanying R script implements the complete data-generating process, the joint EM algorithm, multiple initializations, convergence diagnostics, label alignment, score-based posterior reconstruction, product-of-marginals analysis, and all recovery metrics reported in Tables~2--9. No confidential, proprietary, or borrower-level bank data are used in this study.

\end{document}